\documentclass[preprint2]{aastex}

\shorttitle{Spectroscopic variability of IRAS\,22272+5435}
\shortauthors{Za\v{c}s et al.}

\begin{document}

\title{Spectroscopic variability of IRAS\,22272+5435}


\author{Laimons Za\v{c}s\altaffilmark{1,2},  Faig Musaev\altaffilmark{3,4,5},
             Bogdan Kaminsky\altaffilmark{6}, Yakiv Pavlenko\altaffilmark{6}, Aija Grankina\altaffilmark{1}, 
            Julius Sperauskas\altaffilmark{2}, and Bruce J. Hrivnak\altaffilmark{7}}

% Viktoras Deveikis\altaffilmark{1},      
%  Vasily Puzin\altaffilmark{5},

\altaffiltext{1}{Laser Center, University of Latvia, Rai\c{n}a bulv\= aris 19, LV-1586~R\=\i ga,Latvia}
\altaffiltext{2}{Vilnius University Observatory, \v{C}iurlionio 29, Vilnius 2009, Lithuania}
\altaffiltext{3}{Special Astrophysical Observatory of the Russian AS, Nizhnij Arkhyz, 369167, Russia}
\altaffiltext{4}{ Institute of Astronomy of the Russian AS, 48 Pyatnitskaya st., 
           119017, Moscow, Russia }
\altaffiltext{5}{ Terskol Branch of Institute of Astronomy of the Russian AS,
           361605 Peak Terskol, Kabardino-Balkaria, Russia}
\altaffiltext{6}{Main Astronomical Observatory of Academy of Sciences of Ukraine,  Zabolotnoho 27, Kyiv, 03680, Ukraine}
\altaffiltext{7}{Department of Physics and Astronomy, Valparaiso University, Valparaiso, IN 46383, USA}

\begin{abstract}
A time series of high-resolution spectra was observed in the optical wavelength region for the bright proto-planetary nebula IRAS\,22272+5435 (HD\,235858), along with a simultaneous monitoring  of its radial velocity and $BVR_C$  magnitudes.  
The object is known to vary in light, color, and velocity due to pulsation with a period of 132 days.
The light and color variations are accompanied by significant changes in spectral features, most of which are identified as lines of carbon-bearing molecules.  According to the observations, the $C_2$ Swan system and CN Red system lines are stronger near the light minimum. A photospheric spectrum of the central star was calculated using new self-consistent atmospheric models. The observed intensity variations in the $C_2$ Swan system and CN Red system lines were found to be much larger than expected if due solely to the temperature variation in the atmosphere of the pulsating star. In addition, the molecular lines are blueshifted relative to the photospheric velocity. The site of formation of the strong molecular features appears to be a cool outflow triggered by the pulsation. The variability in atomic lines seems to be mostly due variations of the effective temperature during the pulsation cycle.  The profiles of strong atomic lines are split, and some of them are variable in a time scale of a week or so, probably because of shock waves in the outer atmosphere.
\end{abstract}


\section{Introduction}

The proto-planetary nebula (PPN) phase is an important but still poorly understood stage in the evolution of low- and intermediate-mass stars. Objects in this stage are in transition between the asymptotic giant branch (AGB) and planetary nebula (PN) phases.  A PPN is characterized by a central star of intermediate temperature (5,000 -- 30,000~K) and low surface gravity surrounded by an expanding circumstellar envelope (CSE) of molecular gas and cool dust \citep{kwok}.  

IRAS\,22272+5435 is a well known PPN associated in the optical region with the bright, carbon-rich peculiar supergiant HD\,235858 of spectral type G5 \citep{hrivnak91}.   
The star is surrounded by the detached CSE, a result of earlier AGB mass loss. The morphology of IRAS\,22272+5435 has been studied at various wavelengths. Mid-IR images at arcsecond and subarcsecond resolutions show an elongated emission core that is interpreted as a result of an inclined dust torus or disk \citep{meixner, dayal, ueta2}. High-resolution optical images obtained by the Hubble Space Telescope ({\it HST}) revealed a reflection nebulosity that is elongated approximately perpendicular to the core elongation seen at the mid-IR images \citep{ueta1}. 
\citet{ueta2} carried out a dust radiative transfer model analysis of IRAS\,22272+5435, and their best-fit model  consists of a central star surrounded by two separate sets of dust shells representing an AGB wind shell and a post-AGB wind shell located in the inner cavity of the PPN shell. 
They concluded that IRAS\,22272+5435 left the AGB about 380 yr ago, after the termination of the superwind, and then experienced post-AGB mass loss with a sudden mass ejection about 10 years ago.
Near-IR observations support post-AGB mass ejection for IRAS\,22272+5435 \citep{hrivnak94}.  \citet{nakashima} modeled the mapping of CO emission in the CO\,(2-1) line by an expanding torus of a constant radial expansion velocity of 7.5~$km\,s^{-1}$, with an inner/outer radius of 0.4/1.0\arcsec, and an expanding sphere  representing the AGB wind with an outer radius of 2.5\arcsec and a maximum expanding velocity of 10.5~$km\,s^{-1}$.
Their observations revealed an additional axisymmetric region possibly created by the interaction between an invisible jet and the ambient material. 

IRAS\,22272+5435 is one of the coolest known PPNe and possesses an excess of carbon and s$-$process elements in the atmosphere, synthesized during evolution on the AGB \citep{hrivnak91, zacs95, zacs99, reddy02}. Complex light variations are observed for IRAS\,22272+5435 (V354 Lac), with a maximum amplitude of $\sim$0.5 mag ({\it V}) and evidence of multiple periods that modulate the light curve \citep{hrivnak}. The dominant period of about 132 days attributed to pulsations was found both in the light, color, and radial velocity curves. Despite the enhanced carbon abundance in the atmosphere of IRAS\,22272+5435, molecular lines should be negligible in the stellar spectrum for a G5\,I star.  However, \citet{hrivnak91} detected prominent absorption features of $C_2$ and $C_3$ molecule at 4737, 5165, and 4050 \AA\ using low-resolution spectra. 
\citep{zacs95} confirmed the presence of the strong $C_2$ Swan system (0,0) bandhead at the photospheric velocity using a high-resolution spectrum observed on August 1992.  Information about the site of formation of molecular lines is scarce in the published papers. Not clear whether the peculiar spectrum of  IRAS\,22272+5435 in comparison with that for a typical G5 supergiant is because of the atmospheric or circumstellar anomalies. \citet{bakker} observed a few high-resolution spectra between 25 July and 24 August 1994 in the wavelength region from about 4\,000 to 10\,000 \AA\, and they detected absorption lines of the $C_2$ (0,0), (1,0) Swan system, the $C_2$ (1,0), (2,0), (3,0) Phillips system, and the CN (1,0), (2,0), (3,0), (4,0) Red system.  Most of them were identified to be of circumstellar origin, with a full width at half maximum (FWHM) of $\sim$ 6.0~$km\,s^{-1}$ and  a blueshift between 8.7 to 9.1 $\pm 2.0~km\,s^{-1}$ relative to the adopted systemic velocity, $RV_{sys}^{\sun}$ = -43.1~$km\,s^{-1}$.  The line forming region of these sharp molecular lines was attributed to the AGB ejecta \citep{bakker}. In addition - CN Red system lines of the photospheric origin were found; these lines are much broader  and belong to higher energy levels.
\citet{reddy02} detected the $C_2$\,(2,0) Phillips system circumstellar lines around 8770 \AA\  using a high-resolution spectrum  observed on 2000 June 14. The expansion velocity of the line forming region was measured, with $v_{exp} = 6.3 \pm$ 0.3~km$\,s^{-1}$ relative to the adopted systemic velocity $RV_{sys}^{\sun}$ = -41.0~$km\,s^{-1}$.  However, \citet{reddy02} did not find CN and $C_2$ lines  of the photospheric origin in the spectrum and concluded that IRAS\,22272+5435 is too warm to show photospheric molecular features. 
The absence of carbon bearing molecular lines of the photospheric origin in the spectrum observed by Reddy et al. (2002)  in comparison with the spectrum investigated by \citet{bakker} was not discussed. Moreover, \citet{reddy02} concluded that they detected circumstellar molecular lines  in the spectrum of IRAS\,22272+5435 for the first time.
\citet{zacs09} identified sharp $C_2$ Phillips system and CN Red system lines of circumstellar origin in four high-resolution spectra observed between November 2002 and February 2008. The expansion velocity of line forming region was measured using $C_2$ Phillips system lines as $v_{exp} =  8.4 \pm$ 0.5~$km\,s^{-1}$,  relative to the adopted systemic velocity of $RV_{sys}^{\sun}$ = -40.2~$km\,s^{-1}$.  Broad and variable lines of  $C_2$\,(0,1)  Swan system and CN\,(5,1)  Red system lines were observed simultaneously in the spectra, but blueshifted relative to the systemic velocity by about 10--25~$km\,s^{-1}$.   The site of formation for these broad molecular lines was hypothesized to be a cool outflow. \citet{schmidt} compared the expansion velocity measured using the $C_2$ Phillips system lines in the spectrum observed on November 2002 by \citet{zacs09} with those observed by \citet{bakker} and found a difference of about 3~$km\,s^{-1}$. A comparison of equivalent widths for selected circumstellar lines of $C_2$ Phillips system revealed differences of up to 50 \%  between the two spectra. The reasons for  such large discrepancies in the measurements are difficult to explain by uncertainties in the analysis of these spectra. 

This paper presents the results of an analysis of the time series of high-resolution spectra for IRAS\,22272+5435  in the optical region. In total, 17 spectra were observed between November 2002 and November 2011, simultaneously with light and radial velocity monitoring. The character of the spectroscopic variability is described. Synthetic spectra calculated with new self-consistent hydrostatic atmospheric models  were fitted in selected wavelength regions to the spectra observed during light maximum and minimum to investigate the  reason for the spectroscopic variability.

\section{Observations and reduction}

\begin{figure*}  
      \resizebox{\hsize}{!}{\includegraphics{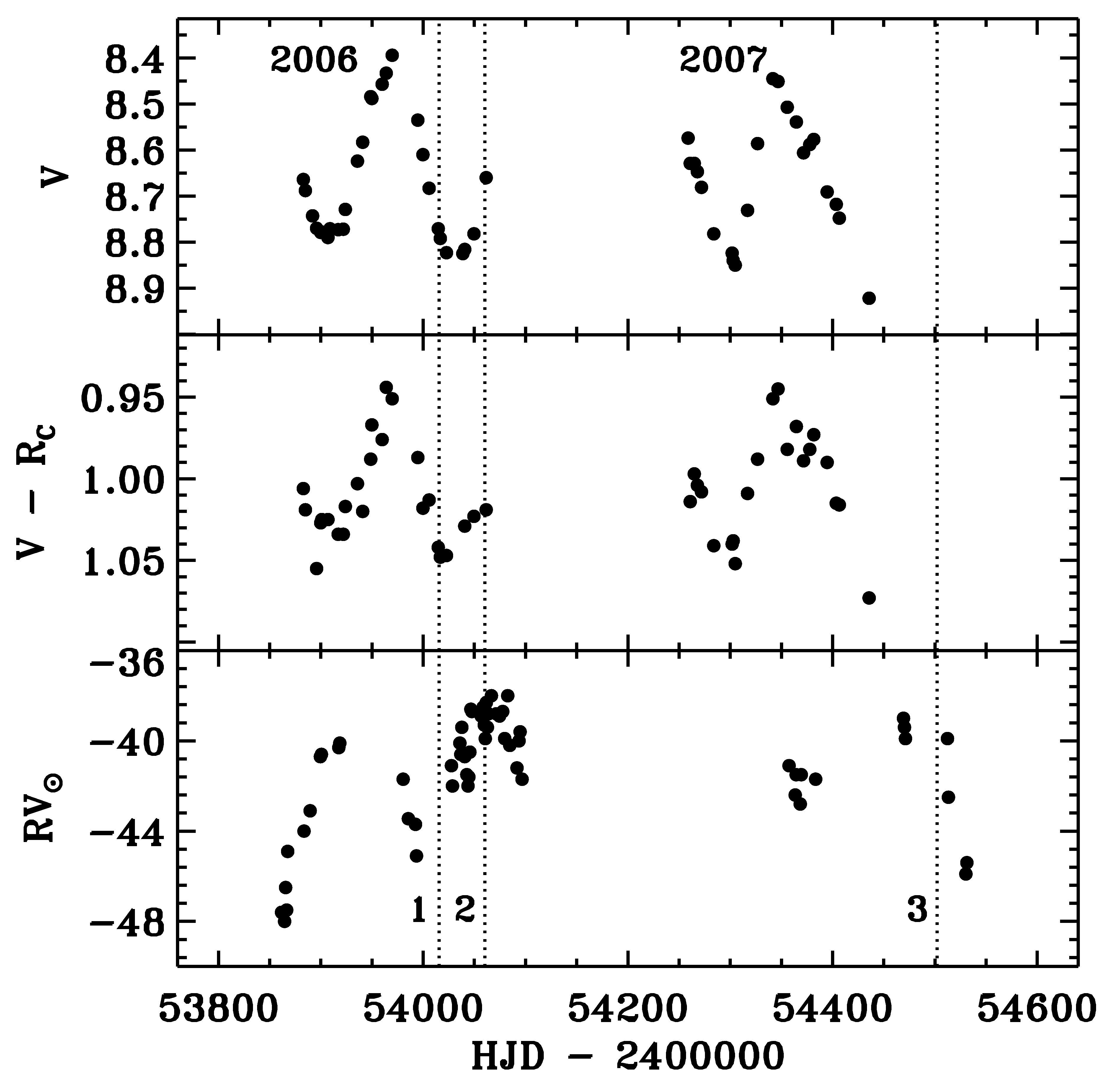}
                             \includegraphics{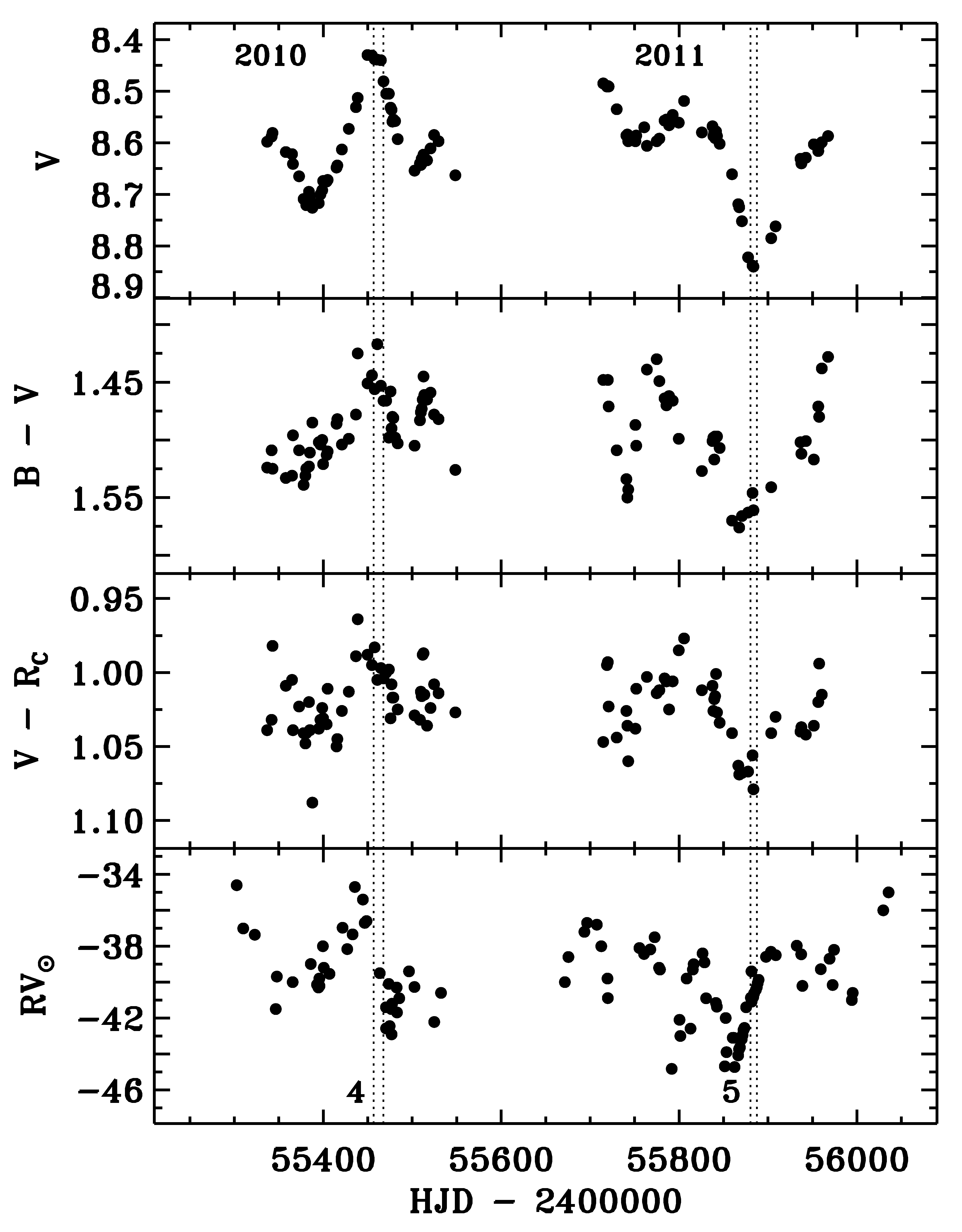}}        
      \caption{Contemporaneous V, (B-V), (V-$R_C$), and $RV_{\sun}$ curves of IRAS\,22272+5435 (HD\,235858) in 2006$-$2007 and 2010$-$2011.  Photometric observations in the standard B filter are absent until 2008.
                    The time (spans) of high-resolution observations are indicated by dotted vertical lines: 1 -- October 2006, 
                    2 -- November 2006, 3 -- February 2008, 4 -- September 2010, 5 -- November 2011.}
      \label{RV}
\end{figure*}

High-resolution spectra of IRAS\,22272+5435 were observed with the coud\'e \'echelle spectrometer MAESTRO on  the 2 m telescope at the Observatory on the Terskol Peak in Northern Caucasus (altitude of 3100 m), equipped with a Wright Instruments CCD detector. Four spectra were initially obtained between 2002 November and 2008 February with the total exposures from 4800 to 7200~s.  In addition, two more extensive sets of observations were performed on September 2010 and November 2011; 7 and 6 high-resolution spectra were observed in the time span of 12 and 8 days, respectively. The spectra cover a wavelength region from about 4000 to 9700 \AA, overlapping blueward of $H_{\alpha}$, and have a resolution R $\sim$ 45\,000. The S/N ratios are of the order of 100 near the \ion{Na}{1} D doublet.  All the spectra were bias subtracted, flat-field corrected, and converted to one-dimensional spectra using the standard DECH20T package.\footnote{http://www.gazinur.com/Spectra-Processing.html}  The wavelength calibration was made using Th$-$Ar and sky spectra obtained for each night. In addition, a spectrum of a hot and rapidly rotating star was observed to identify the telluric absorption lines. Details of the spectroscopic observations are included in Table~\ref{log}.  The phases were calculated using the 132 days period where phase 0.0 is defined by the time of the nearest light minimum. The epochs of the high-resolution observations relative to  the light, color, and radial velocity curves published by \citet{hrivnak} are shown in Figure~\ref{RV}.  

The equivalent widths (EW) of the absorption lines were measured by Gaussian fitting to the observed profiles. The radial velocities were estimated using the standard DECH20T routine: a direct and mirror profile of each selected line was correlated and the shift of the central wavelength from its rest wavelength was measured. For a sample of measured lines, the mean radial velocity and its standard deviation was calculated to characterize the random error. The heliocentric corrections were calculated using the standard DECH20T routine. We adopted the systemic heliocentric velocity  $RV_{sys}^{\sun}$ = -40.8~$km\,s^{-1}$ calculated on the basis of radial velocity data collected for IRAS\,22272+5435  in the time span from 1988 to 2011 \citep{hrivnak}, which is close to the systemic velocity measured using CO emission lines of the CSE, $RV_{sys}^{\sun}$ = -40.2~$km\,s^{-1}$  \citep{hrivnak05}. The systematic errors of the radial velocity measurements were examined by measurements of wavelengths for telluric lines identified in the stellar spectra. A typical standard deviation for a sample of measured weak and medium strong symmetrical lines is about 0.5~$km\,s^{-1}$  and the systematic errors are below 1.0~$km\,s^{-1}$.  The measured instrumental photospheric radial velocities, heliocentric corrections, heliocentric photospheric radial velocities, and instrumental systemic radial velocities are given in Table~\ref{T_RV}.

\begin{table}
\caption{The log of high-resolution spectroscopic observations. Times of observations, exposures and 
approximate phases of light variation are given. \label{log}}
\vskip 0.1in
\begin{tabular}{rcrc}
\hline \hline
Time   &  HJD  &   Exp   &  Phase\tablenotemark{a}          \\
dd/mm/yy  & (-2450000)  &  (min)   &    \\
\hline  
18.11.2002 &   2597.3    & 120         &  0.5\tablenotemark{b}  \\  
 6.10.2006 &  4015.4    &90            &   0.9 \\
20.11.2006 &  4060.3     &80           &    0.2  \\
 5.02.2008 &   4502.2    &120          &   0.5\tablenotemark{b}   \\
17.09.2010 & 5456.5     &90           &  0.5       \\
20.09.2010  &  5459.5    &90        &  0.5   \\
21.09.2010  & 5460.5     & 120       &  0.5   \\
22.09.2010 &  5461.5    & 90        &   0.5  \\
25.09.2010 &  5464.5     & 90    &   0.5  \\
27.09.2010 &   5466.5    &  90      &   0.5  \\
28.09.2010 & 5467.5      & 120       &   0.5 \\
14.11.2011 & 5880.4       &   90      &  0.0  \\
15.11.2011 &  5881.3       &  90      &  0.0  \\
18.11.2011 &  5884.3       &   90    &    0.0 \\
19.11.2011 &   5885.4      &  90     &   0.0  \\
20.11.2011 &  5886.3      &   120    &   0.0  \\
21.11.2011 & 5887.4       &  120     &   0.0  \\
\hline
\end{tabular}
\tablenotetext{a}{Phase based on P = 132 days, where phase = 0.0 is defined by\\
 the time of nearest light minimum.}
\tablenotetext{b}{Phase based on the interpolation/extrapolation of the light \\ curve by a sine  wave with P = 132 days.}
\end{table}

\begin{table}
\caption{Radial velocities measured using a sample of weak and medium strong atomic lines in the high resolution spectra.
Time of observation, instrumental photospheric radial velocities, heliocentric corrections, heliocentric photospheric radial velocities, and instrumental systemic radial velocities are given. \label{T_RV}}
\vskip 0.1in
\begin{tabular}{rcrcc}
\hline   \hline
Time   &  RV &  $V_a$  &  $RV_{\sun}$ &  $RV_{sys}$    \\  
       &    (km\,s$^{-1}$)    &     &    &         \\
\hline  
18.11.02 & -33.1 & -11.54 & -44.6 & -29.3    \\  
6.10.06  &  -42.5 & -0.57 & -43.1 &  -40.2   \\
20.11.06 &  -25.1 & -11.99 &  -37.1 &  -28.8   \\
5.02.08 & -25.7 & -14.04 & -39.7 & -26.8  \\
22.09.10 & -45.8 & +3.47 & -43.1 & -44.3   \\
18.11.11 & -28.6 & -11.53 &  -40.1 & -29.3   \\   
\hline
\end{tabular}
\end{table}

\section{Analysis and modeling}

\subsection {Description and comparison of the spectra}

\begin{figure*}
        \resizebox{\hsize}{!}{\includegraphics{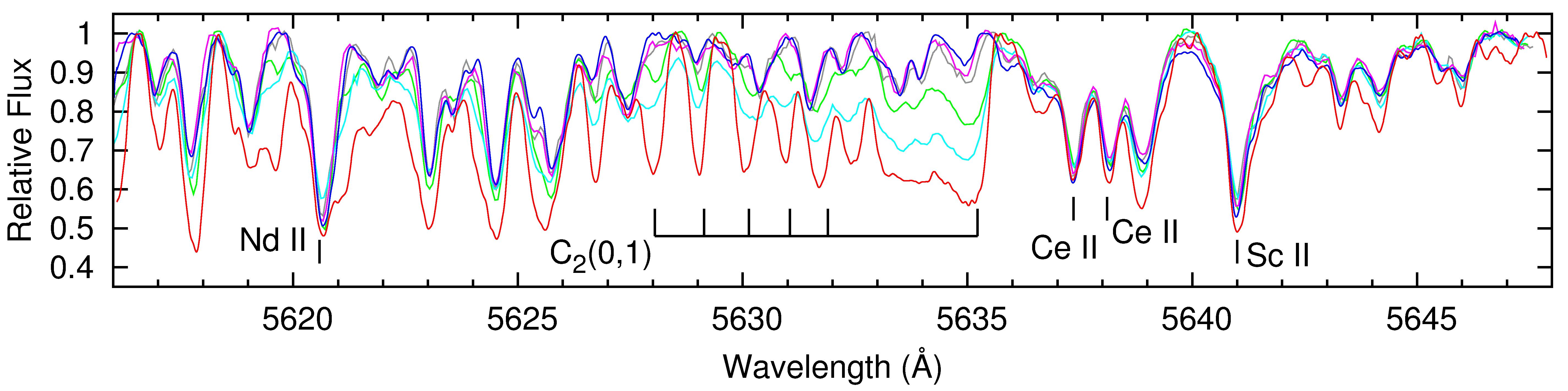}}
    \caption{IRAS\,22272+5435 spectra in the wavelength region from 5616 to 5648 \AA\ observed on November 2002 (grey; phase = 0.5), October 2006 (cyan; phase = 0.9), November 2006 (green; phase = 0.2), February 2008 (magenta; phase = 0.5),  September 2010 (blue; phase = 0.5), and November 2011 (red; phase = 0.0). The wavelengths are corrected for the instrumental photospheric radial velocities given in Table~\ref{T_RV}.  The $C_2$ Swan system (0,1) bandhead and some of less blended lines are marked.}
    \label{C2_5635_5}
\end{figure*}

\begin{figure*}
       \resizebox{\hsize}{!}{\includegraphics{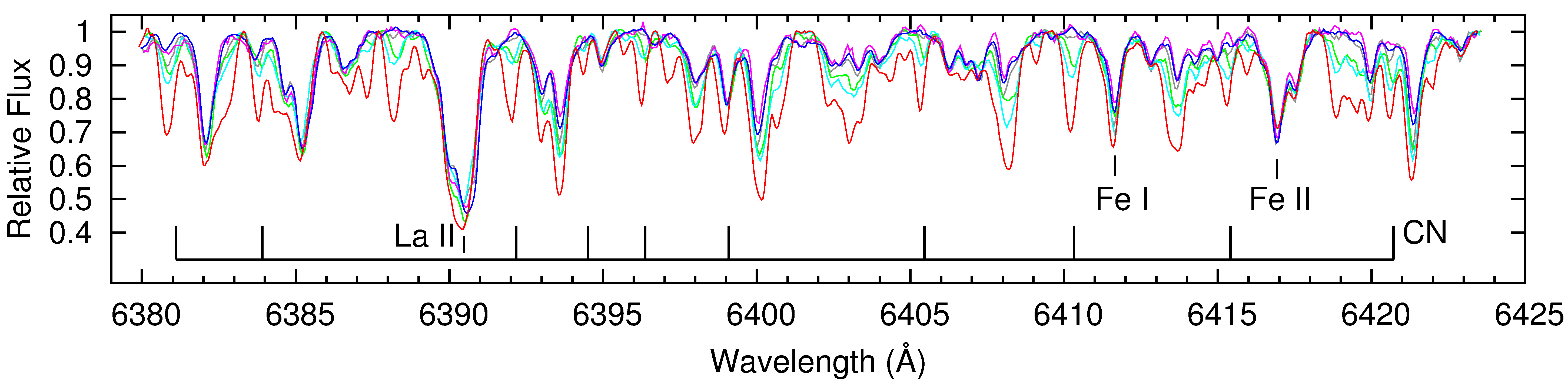}}
    \caption{Same as Fig.~\ref{C2_5635_5}, but in the region from 6380 to 6423 \AA. CN (5,1) Red system lines  and two less blended iron lines are marked. A broad and variable absorption feature at 6390.5 \AA\ formed mostly by the  \ion{La}{2}  line is shown. Weak emissions are seen in CN lines near light maximum.}
    \label{CN_51_5}
\end{figure*}

\begin{figure}
 \resizebox{\hsize}{!}{\includegraphics{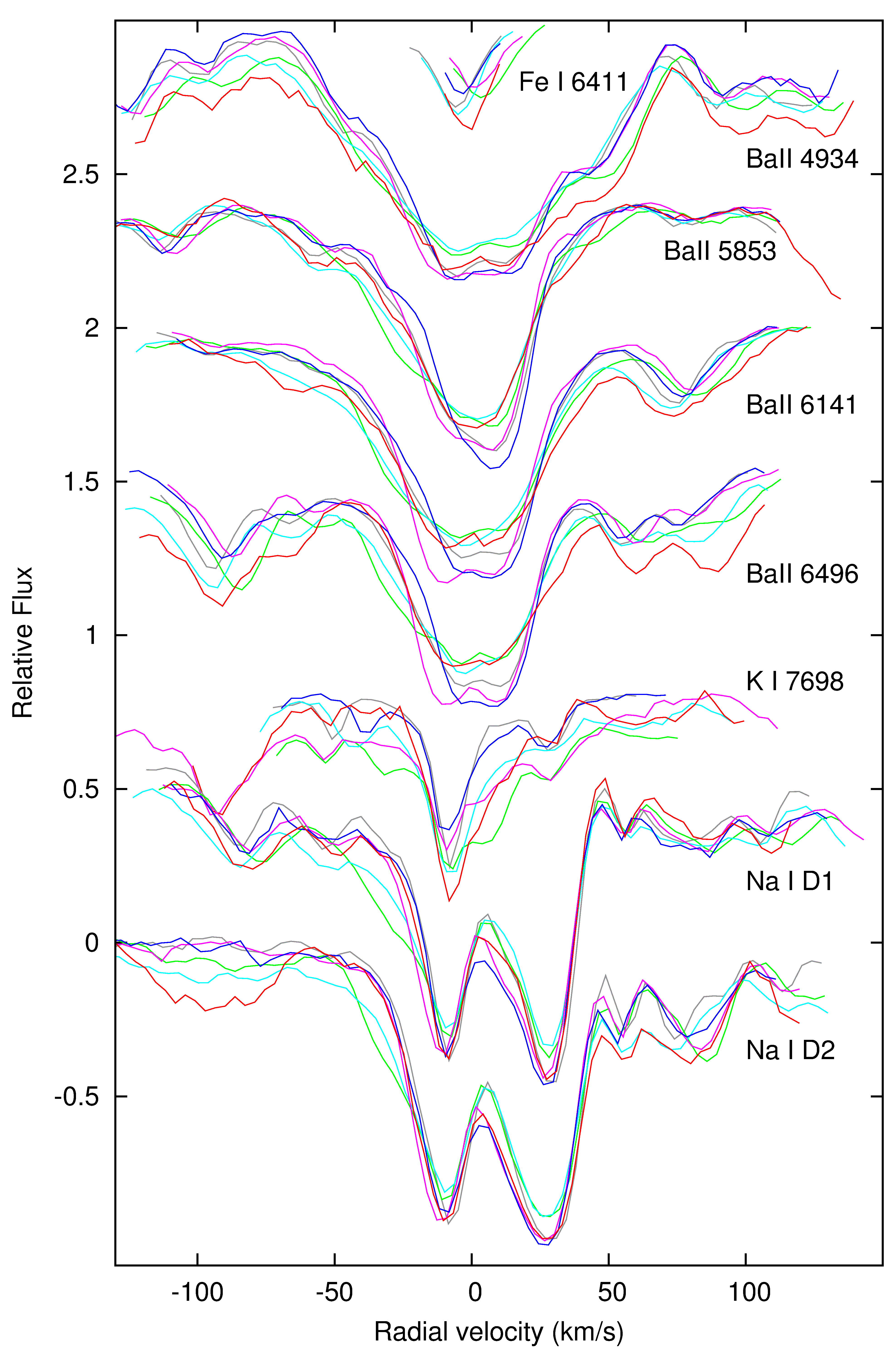}}
 \caption{Profiles of seven strong low-excitation lines observed on November 2002 (grey), October 2006 (cyan), November 2006 (green), February 2008 (magenta), September 2010 (blue), and November 2011 (red) in the radial velocity scale relative to the systemic velocity. A medium strong photospheric \ion{Fe}{1} line at 6411.65 \AA\ (EW $\sim$ 100 m\AA; LEP = 3.65 eV) is shown for comparison purposes.}
   \label{split}
\end{figure}

The optical spectrum of IRAS\,22272+5435 is more complicated than is generally seen in stars of approximately solar temperature.  
A number of lines of heavy elements from weak transitions  are visible which are due to the strong enhancement of neutron-capture  elements, and these s-process lines are significantly blended with the lines of another species throughout the spectrum. In addition, a numerous weak and sharp $C_2$ and CN lines are present in the spectrum originated probably in the AGB shell.  Moreover, broader in comparison with the shell lines and variable features  of carbon bearing molecules are visible  in some of the observed spectra (see Figures~\ref{C2_5635_5} and ~\ref{CN_51_5}). We inspected a number of lines from the blue to near-infrared wavelength region and concluded that  most of the absorption features are variable in a timescale of a month to a few years.  A smaller number of spectral features seem to be variable in a timescale of a week. A correlation of the intensity of the $C_2$ bandheads with the phase of the light variation is clearly visible in the spectra - the Swan system bandheads are weak near light maximums (e.g. on September 2010 and  February 2008) and strong near light minimums (e.g. on November 2011 and October 2006). 
A shift of wavelengths is clearly seen for the $C_2$ bandheads relative to the photospheric radial velocity, the later of which was measured on the basis of weak and symmetrical atomic lines. The intensity variation of CN Red system lines in general follows the intensity variation of  $C_2$ Swan system lines (see  Figure~\ref{CN_51_5}). In addition, the CN (5,1) Red system lines seem display weak emission near the observed light maximums. The profiles of strong atomic lines are broad and split into two or three components that are partly resolved at this resolution (Figure~\ref{split}). Comparison of seven spectra observed during the set of observations on September 2010 revealed only minor spectroscopic variability, which is seen only in some strong lines (see Section 3.4).  Therefore, we combined these spectra, observed in a time span of 12 days, with the goal of increasing the S/N ratio (especially in blue wavelength region). A similar procedure was performed for the six spectra gathered during the set of observations on November 2011. According to the photometry (see Figure~\ref{RV}), IRAS\,22272+5435 is at its light maximum on September 2010 and at its light minimum on November 2011.  A quantitative analysis of the spectroscopic variability was performed using the combined spectra observed at these two light extrema.

\subsection{Atmospheric parameters and abundances}

The value of effective temperature for the central star of  IRAS\,22272+5435  was determined from the constraint that the relation between iron abundances (from individual \ion{Fe}{1} lines) and the excitation potential have a zero slope, resulting in $T_{eff}$ = 5600 $\pm$ 250~K \citep{zacs95}. \citet{reddy02}, using the same method, obtained a similar value within the uncertainty of $T_{eff}$ = 5750~K. The best-fit model of spectral energy distribution (SED) for IRAS\,22272+5435 gives $T_{eff} \sim$ 5800~K for the central star \citep{ueta2}.  The interstellar extinction was found to be $A_V$ = 2.5 mag for this best-fit model, which is larger than the observed value along the line of sight towards IRAS\,22272+5435 for the adopted distance of 1.6 kpc \citep{neckel}. The determination of the effective temperature on the basis of color indices entails many complicated problems for the post-AGB stars. Chemical peculiarity and the presence of dust are responsible for the continuum depression in the blue and excess in the infrared wavelength regions,  both affecting the observed colors. The central star of IRAS\,22272+5435 is a pulsating star with time-varying color indices.  This pulsation leads to temperature variation in the photosphere of the star where most of the absorption lines are formed. Thus these changes of $T_{eff}$ during the pulsation cycle should contribute to the observed spectroscopic variability. 
A crude estimation of the temperature variation for the central star was performed using the color index $(V-R_C)$, which is less affected by  the continuum anomalies than is $(B-V)$. A maximal variation of the color temperature was estimated using the $(V-R_C)$ observed  at the light maximum on September 2010 and that at the light minimum on November 2011, when two sets of  high-resolution spectroscopic observations were gathered. Assuming a color temperature relationship for IRAS\,22272+5435 similar to that for  a typical supergiant, the color variation of $\delta (V-R_C)  \simeq$ 0.12 mag translates into the temperature variation of about 500~K for the star with $T_{eff}$ = 5750~K \citep{johnson}. Bearing in mind the published values of effective temperature for IRAS\,22272+5435 and the probable temperature variation in the atmosphere due to pulsation, two final temperatures were adopted for the modeling of spectrum for the central star, $T_\mathrm{eff}$ = 5750  and 5250~$\mathrm{K}$ (the hotter and cooler atmospheric models, respectively).

The surface gravity was adopted to be $\log g$ = 0.5 (cgs) for the final models, which is a typical value for the  cool post-AGB stars \citep{pereira} and agrees well with the gravity determined for IRAS\,22272+5435 using a ionization balance for iron.  In addition, we examined some  atmospheric models with lower $T_\mathrm{eff}$ and the gravity in a range from $\log g$ = 0.0 to 1.5 (cgs). According to \citet{pereira}, $\log g$ = 0.5 and 0.2 (cgs) on average for post-AGB stars with the $T_\mathrm{eff}$ = 5750  and 5250~$\mathrm{K}$, respectively (see Figure 2 in the cited paper). The microturbulent velocity was adopted to be $\xi_t$ = 4.5 km\,s$^{-1}$, which is close to the mean published value \citep{zacs95, zacs99, reddy02} found by forcing the calculated abundances from individual iron lines to be independent of the equivalent widths.  \citet{zacs99} found evidence that the value of the estimated microturbulent velocity for IRAS\,22272+5435 depends on the upper limit of the used EWs, with a tendency to give  larger $\xi_t$ for stronger lines. 

\begin{table*}
\begin{center}
\caption{The atmospheric parameters and abundances collected from literature.
\label{basic_dat}} 
\begin{tabular}{ccccccccc}
\tableline\tableline
 $T_\mathrm{eff}$  & $\log g$ & $\xi_t$ & [Fe/H] & [C/H]  & [N/H]  & [O/H] & $[n/H]^a$  & Ref. \\
   (K)    &  (cgs)  & (km\,s$^{-1}$)  &  (dex) & (dex) & (dex) & (dex) & (dex) & \\
\tableline
5600  &  0.5 & 3.7  & -0.49  & +1.19  & ...  &  -0.10  &  +1.9 (6) & 1   \\
5600  & 0.5  & 3.7, 7.0 & -0.5  &+0.8  &  ... & -0.2 &  +1.5 (17) & 2 \\
5750 & 0.5 & 4.5  & -0.82  &+0.17   & -0.24 & -0.30  & +1.2 (7) & 3  \\
5800  & ...  & ...    &  ...& ...   & ...   &  ... &  ... &  4  \\
\tableline
\end{tabular}
\tablenotetext{a}{The average enhancement of neutron capture elements. The number of species is given in parentheses.}
\tablerefs{(1)~ \citet{zacs95}; (2)~ \citet{zacs99}; 
 (3)~\citet{reddy02}; (4)~\citet{ueta2}}
\end{center}
\end{table*}

We adopted the abundances of CNO elements calculated by \citet{reddy02}, [C/H] =  0.17 dex, [N/H] = --0.24 dex, and [O/H] = --0.30 dex, which lead to the carbon-to-oxygen ratio C/O = 1.6.  The standard notations are adopted everywhere.\footnote{[A/B] = $\log\,(N_A/N_B)_{\star}$ - $\log\,(N_A/N_B)_{\sun}$, where $N_A$, $N_B$  is the number density of an element A and B. \\ $\log\,\epsilon(A)$ = $\log\,(N_A/N_H)$ + 12.00, where $N_H$ is the number density of hydrogen.}  A medium iron deficiency, [Fe/H] = --0.7 dex, and  an enhanced abundances for all neutron capture elements were adopted, [n/H] = +1.5 dex. We adopted the scaled solar abundances, [X/Fe] = 0.0, for the rest elements, despite of some overabundances reported for IRAS\,22272+5435 in the literature. The fits between the synthesized atomic spectra and the observed ones confirm  in general the correctness of the adopted atmospheric parameters and chemistry (see Sections 3.4, 3.5, and 3.6), although for some species we detected the difference over a typical value of 0.2 dex in comparison with the abundances calculated by \citet{reddy02}. The atmospheric parameters and abundances collected from literature are given in Table~\ref{basic_dat}. 
Updates of the published abundances are foreseen using the time series of spectra and the new models in a future paper. On the other hand, the optical spectrum of IRAS\,22272+5435 originates at various depths in the atmosphere of central star and these are affected differently by pulsations.  Due to this, a  hydrostatic atmospheric model is not able to describe correctly the formation of all lines. According to the recent studies, the dynamical models fit the observed spectra of long-period variables much better than any hydrostatic model.  However, for some spectral features, the variations in the line intensities predicted by dynamical models over a pulsation cycle give values similar to that of a sequence of hydrostatic models with varying temperature and constant surface gravity \citep{lebzelter}. It is expected that absorption lines that are formed deep in the atmosphere of  low-amplitude pulsating stars are calculated correctly using hydrostatic models. \citet{loidl} demonstrated that hydrostatic atmospheric models are able to reproduce quite well the observed spectra of cool carbon-rich semiregular variables in the large wavelength region from 0.7 to 2.5 $\micron$.

\subsection{Atmospheric models and spectral synthesis}

The standard atmospheric models prepared by \citet{kurucz93} were used for the first iteration of the spectrum synthesis to examine the published abundances. Then a grid of new self-consistent atmospheric models  were calculated in the range of effective temperature  and surface gravity suitable for the central star of IRAS\,22272+5435 (Table~\ref{models}) to check the impact of  models on the calculated  spectrum. The new atmospheric models  were calculated using the code SAM12 (Pavlenko 2003), a modification of the Kurucz code ATLAS12  \citep{kurucz05}, designed to calculate atmospheric models of red giants of a given chemical composition. The modifications are as follows. The bound$-$free absorption caused by the C\,{\sc i}, N\,{\sc i}, and O\,{\sc i} atoms was added \citep{pz03} to the continuum absorption sources included in ATLAS12. We also take into account collision induced absorption -- absorption of the molecular complexes He$-H_2$ and $H_2-H_2$ induced by collisions -- which becomes an important source of opacity in the atmospheres of cool metal-poor stars \citep{borysow}. Molecular and atomic absorption in the SAM12 code are taken into account by using the opacity sampling technique \citep{sneden}. A compiled list of spectral lines includes atomic lines from the VALD3 database \citep{VALD1, kupka} and molecular lines of CN, C$_2$, CO, SiH, MgH, NH, OH from the Kurucz database \citep{kurucz93}. In addition, the absorption by the HCN bands was taken into account according to  \citet{harris, harris2}. We also took into account the isomers HCN and NHC absorption. SAM12 was used successfuly for calculations of the atmospheric models for chemically peculiar stars like R~CrB stars, Sakurai's object, and other evolved stars.

\begin{figure}
 \resizebox{\hsize}{!}{\includegraphics{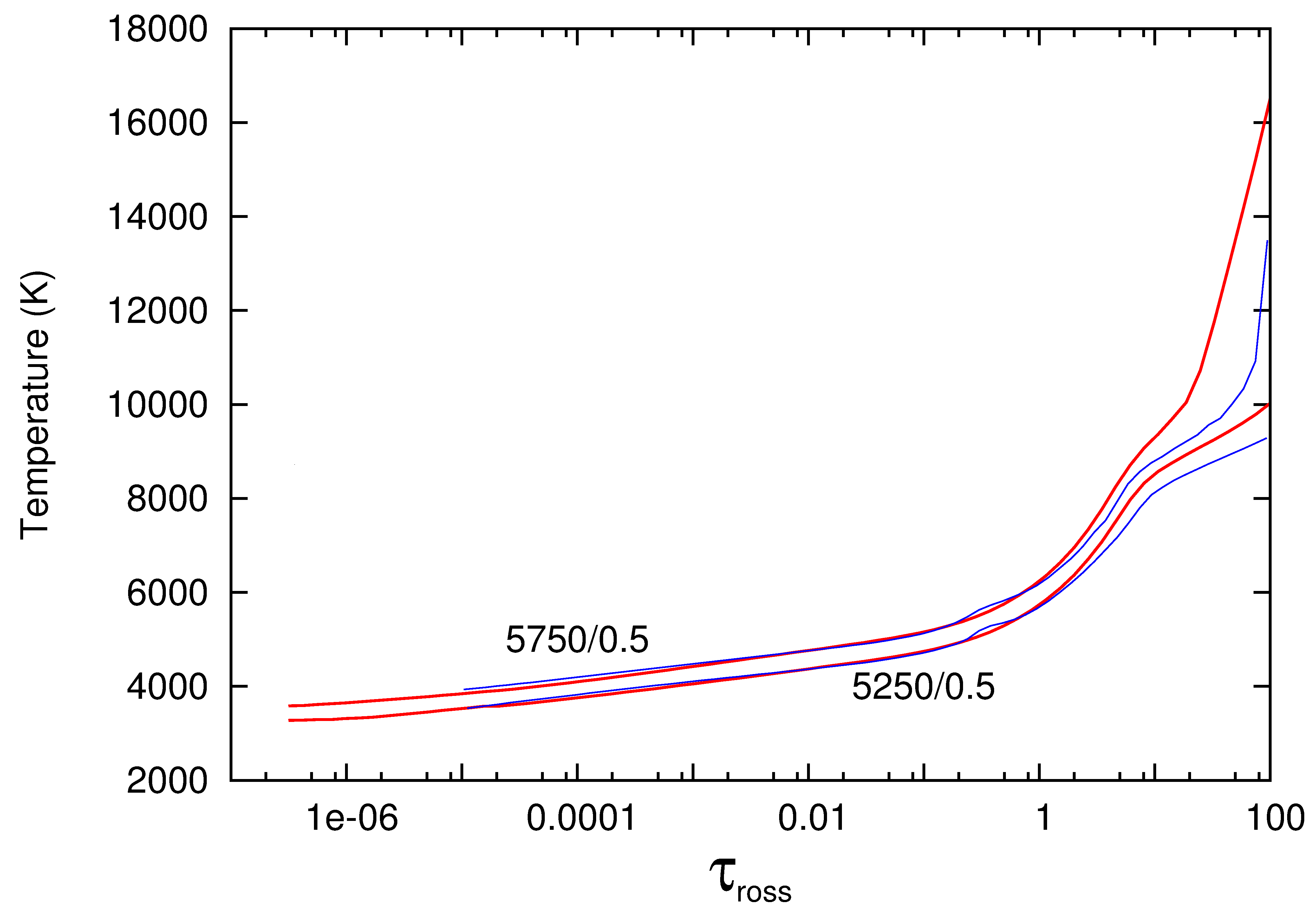}}
 \caption{Temperature structure for two SAM12 atmospheric models  (blue) calculated for  $\log g$ = 0.5 (cgs) and two effective temperatures, $T_{eff}$ = 5750 and 5250~K, along with that for the corresponding \citet{kurucz93} models (red).}
   \label{mod_structure}
\end{figure}

Structure of the calculated atmospheric models depends on the adopted chemical composition, therefore, we used the actual abundances for IRAS\,22272+5435 retrieved from the literature (see Section 3.2). The published metallicities for IRAS\,22272+5435 are in the range of 0.3 dex and the mean weighted metallicity was adopted for calculation, [Fe/H] = $-$0.7 dex. We adopted the carbon and oxygen abundances estimated by \citet{reddy02} because of higher resolution of their spectrum (lower blending effects) and larger number of employed lines in comparison with the previous papers. The published abundances for neutron capture elements are contradictory - the differences between estimations for some elements are much larger than the predicted uncertainties \citep{zacs95, zacs99, reddy02}. The mean enhancement calculated on the basis of seventeen neutron capture elements, [n/H] = +1.5 dex,  was adopted for all the n$-$capture elements in the SAM12 models. 
A comparison of  temperature structure of two SAM12 atmospheric models and the corresponding  Kurucz models is given in Figure~\ref{mod_structure}. As can be seen, the models are similar; however, the outer layers for the SAM12 models are cooler in comparison with the Kurucz models. The synthetic spectra  were calculated in selected wavelength regions with the code WITA \citep{pavlenko97} and were appropriately convolved with a Gausian profile with FWHM  ranging from about 0.4 (blue) to 0.8 (near-infrared) \AA\  depending on the wavelength region. The VALD3 database was used as a primary source of atomic and molecular data for synthesis of the spectra. The list of molecular lines was adopted from the Kurucz database (Kurucz 1993) and SCAN tape \citep{jorgensen}.

\begin{table}
\caption[]{The grid of SAM12 atmospheric models calculated for modeling the spectra. The adopted atmospheric parameters and abundances are  given. \label{models}}
\begin{tabular}{crccc}
\hline
\hline
\noalign{\smallskip}
 $T_{eff}$ & $\log$ g &  [M/H]  & C/O &  [n/H]   \\
  (K)           &   (cgs)     &  (dex)  &        &  (dex)   \\
\noalign{\smallskip}
\hline
\noalign{\smallskip}
  5750  &    1.5  &  -0.7  & 1.6 & +1.5  \\
  5750  &    0.5  &  -0.7  & 1.6 & +1.5  \\
  5750  &    0.1  &  -0.7  & 1.6 & +1.5  \\
  5250  &    0.5  &  -0.7  & 1.6 & +1.5  \\
  5250  &    0.0  &  -0.7  & 1.6 & +1.5  \\
  4750  &    0.5  &  -0.7  & 1.6 & +1.5  \\
  4750  &    0.0  &  -0.7  & 1.6 & +1.5  \\
 \noalign{\smallskip}
\hline
\end{tabular}
\end{table}

\subsection{Atomic lines}

\begin{figure*}
   \resizebox{\hsize}{!}{\includegraphics{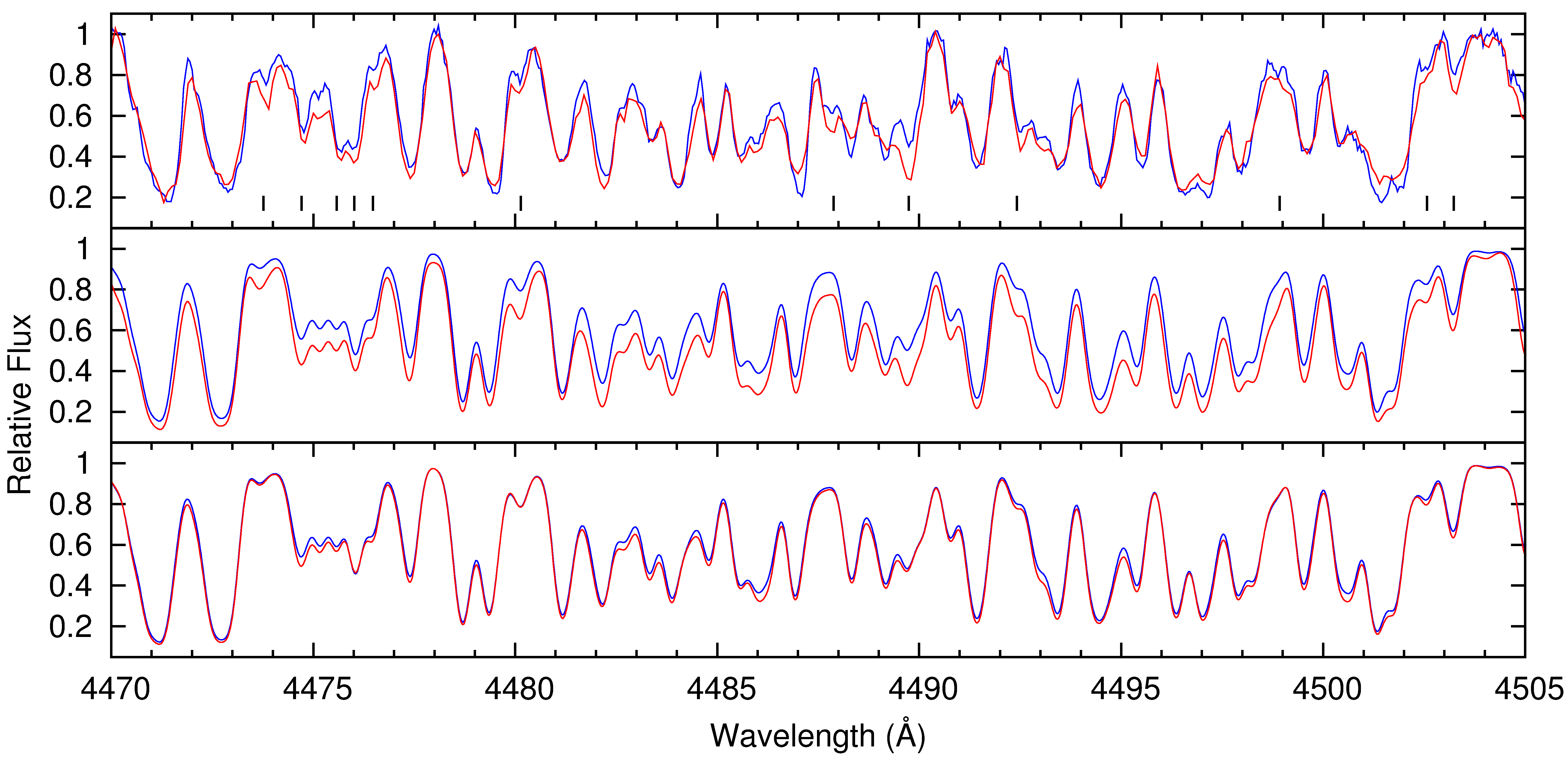}}
    \caption{The combined spectrum observed at light maximum on September 2010 (blue), along with that observed at light minimum on November 2011 (red) in the  wavelength region from 4470 to 4505 \AA\  (upper panel). Some of most variable lines are marked by vertical ticks. The model spectra calculated using SAM12 atmospheric models for  $\log g$ = 0.5 (cgs) and  two effective temperatures, $T_\mathrm{eff}= 5750~ \mathrm{K}$ (blue) and 5250$~\mathrm{K}$ (red), are given in the middle panel. The model spectra calculated for  $T_\mathrm{eff}= 5750~ \mathrm{K}$ and two gravities,  $\log g$ = 0.5 (blue) and 0.1 (red), are given in the bottom panel.}
   \label{atomic_blue}
\end{figure*}

Comparison of the observed spectra shows that most of the absorption lines are variable. However, the largest variations are seen mainly in the molecular lines (see Section 3.5 and 3.6).  Numerous lines of carbon-bearing molecules of different strengths are blended with atomic lines throughout the spectrum, especially near the light minima,  mimicking a variability in the blended atomic lines. On the other hand, the atomic lines could be intrinsically variable as well because of variations of the atmospheric parameters for the central star during the pulsation cycles. The intrinsic variability of the atomic lines was examined in the blue wavelength region, where contamination from molecular lines  is lower in the observed spectra during all phases of the light cycle. An example of the comparison of the spectra observed at light maximum on September 2010 and that observed at  light minimum on November 2011 is shown in Figure~\ref{atomic_blue}. The color curve is also at its extreme values on these two dates, being bluer (hotter) when brighter. As can be seen, most of the spectral features reveal some minor variability.  However, some of them display large intensity variations between the light and color extrema. The most variable atomic lines are marked by tick marks in Figure~\ref{atomic_blue} and mainly result from neutral low-excitation transitions (Table~\ref{atomic_var}). The variability of the spectrum due to temperature variation was modeled using the hot and cool atmospheric models calculated for IRAS\,22272+5435. Synthesized spectra for the final atmospheric models with $T_{eff}$ = 5750~K and 5250~K are displayed in Figure~\ref{atomic_blue} (middle panel). The sensitivity to temperature variation is clearly seen in a number of atomic lines e.g. \ion{Ce}{2} at 4473.77 \AA, \ion{Sm}{2} and \ion{Ce}{2}  at 4476.5 \AA,  \ion{Fe}{1}  at 4489.74 \AA, \ion{Pr}{2}  at 4492.42 \AA, and \ion{Dy}{2}  at 4503.23.  The sensitivity of the selected  lines to the changes of surface gravity was found to be relatively low (see Figure~\ref{atomic_blue}; bottom panel). We conclude that most of the intrinsic intensity variations in the weak and medium strong atomic lines throughout the spectrum of IRAS\,22272+5435 are because of temperature variations rather than to changes in surface gravity in the line forming region during the pulsations cycle.

\begin{table}
\caption[]{Lines in blue wavelength region displaying a significant intensity variation between spectra observed at light maximum and minimum. Species identifications, rest wavelengths, lower  excitation potentials, and oscillator strengths as adopted from the VALD3 database are given.  \label{atomic_var}}
\begin{tabular}{ccccc}
\hline
\hline
\noalign{\smallskip}
 Species  & Wavelength &  LEP   &   $\log gf$ &  Ref.   \\
               & ( \AA\ )        &  (eV)  &                  &           \\
\noalign{\smallskip}
\hline
\noalign{\smallskip}
\ion{Ce}{2}   & 4473.77    &  1.48   & -0.489   & PQWB \\
\ion{V}{1}   &  4474.71   &  1.89   &  0.037  &  K09 \\
\ion{Nd}{2}  & 4475.58   &  0.06   &  -1.827  &  MC \\
\ion{Fe}{1}  & 4476.02   & 2.84    &  -0.819  & K07 \\
\ion{Fe}{1}   & 4476.07   & 3.69    &  -0.175 &   K07\\
\ion{Sm}{2}  & 4476.48   & 0.38   &  -1.759  &   XSQG\\
\ion{Ce}{2}  & 4476.51   & 0.74    & -1.860  &  PQWB \\
\ion{Fe}{1}   &  4480.14  & 3.05    & -1.932  &  BWL\\
\ion{Ce}{2}  & 4487.88   & 1.49    &  -0.789  &  PQWB \\
\ion{Fe}{1}   &  4489.74  &  0.12   & -3.966  &  K07 \\
\ion{Pr}{2}   & 4492.42   &  0.42   &  -1.153  &  MC  \\
\ion{Ce}{2}  & 4498.92   & 1.32    &  -1.349  &  PQWB  \\
\ion{Ce}{2}   & 4502.57   & 1.25    &  -1.250  &  PQWB \\
\ion{Dy}{2}   & 4503.23   &  0.93   &  -1.487  &  MC  \\
\noalign{\smallskip}
\hline
\end{tabular}
\tablerefs{Ref. according to the VALD database:  \\ http://www.astro.uu.se/valdwiki/VALD3linelists}
\end{table}

\begin{figure*}
       \resizebox{\hsize}{!}{\includegraphics{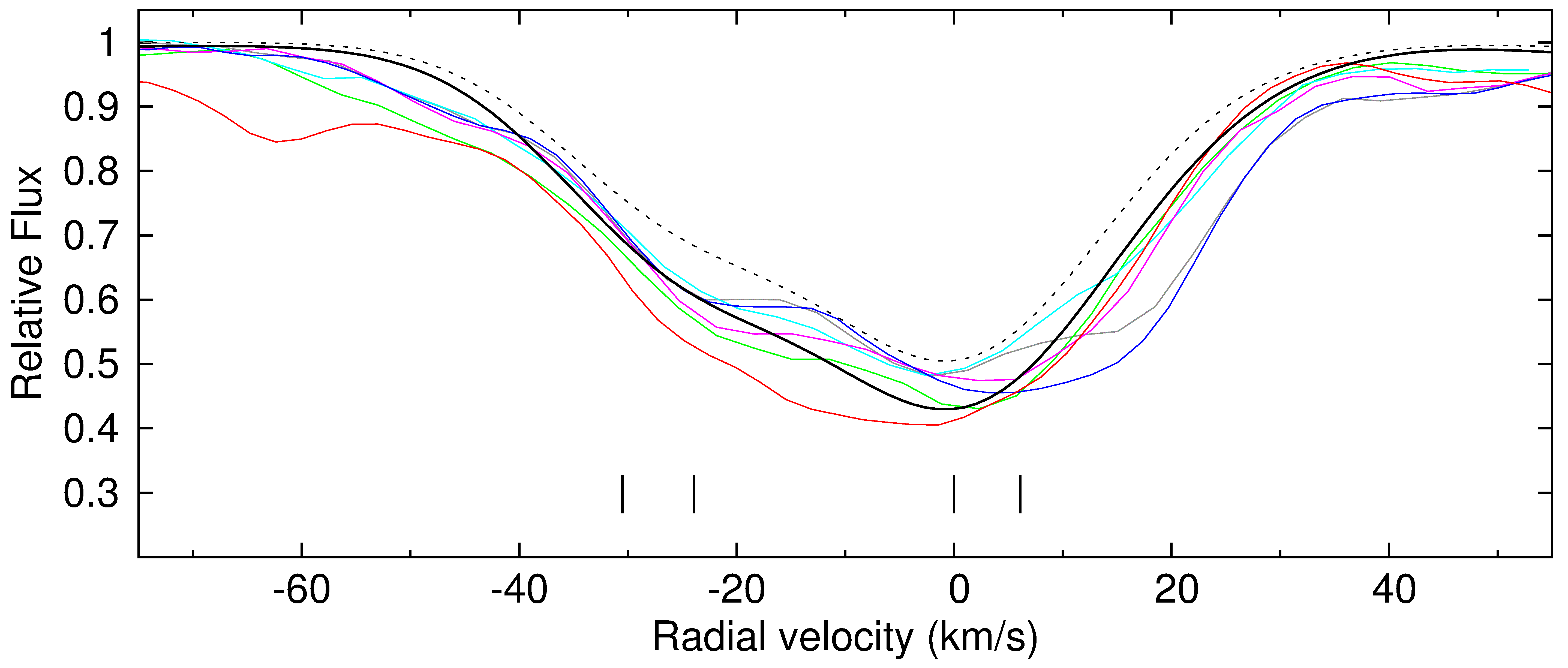}
                             \includegraphics{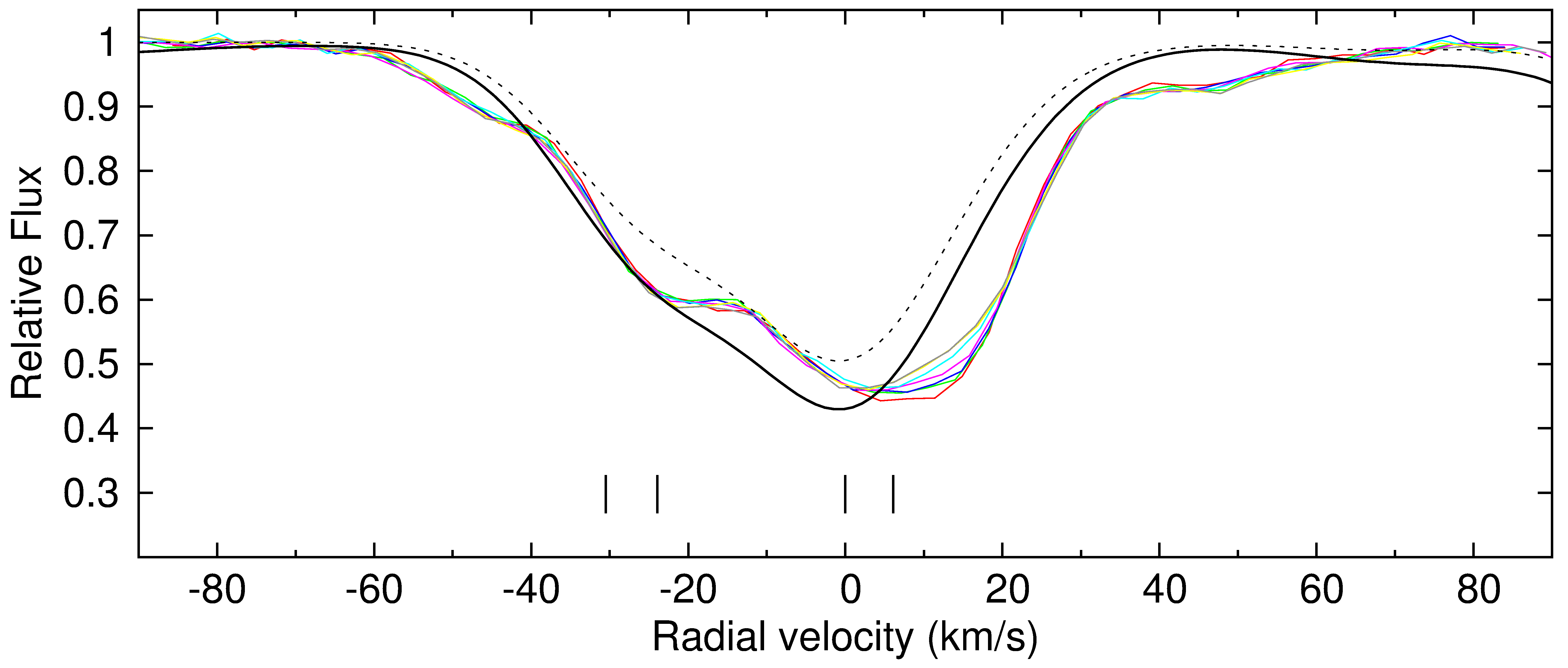}}           
    \caption{Profile of the \ion{La}{2} line at 6390 \AA\ in radial velocity scale relative to the photospheric velocity. Left panel: Observations made at six different epochs: November 2002 (grey), October 2006 (cyan), November 2006 (green), February 2008 (magenta), September 2010 (blue), and November 2011 (red).  
 Right panel: A time series of spectra taken on 2010 September 17 (red), 20 (green), 21 (blue), 22 (magenta), 25 (cyan), 27 (grey), and 28 (yelow). The positions of four contributing atomic lines (including three weaker in comparison with the lanthanum) are marked. The profiles calculated with the SAM12 atmospheric models for  $\log g$ = 0.5 (cgs) and  for two effective temperatures, $T_\mathrm{eff}= 5750~ \mathrm{K}$ (black dashed) and 5250$~\mathrm{K}$ (black solid), are shown. The synthesized profiles are not able to reproduce the observed blue/red-shifted components.}
     \label{La2_profile}
\end{figure*}

Inspection of the observed spectra gives evidence that some of lines are extremely broad and split. The profiles of four \ion{Ba}{2} lines at 4934, 5853, 6141, and 6496 \AA, \ion{K}{1} line at 7698, and the  \ion{Na}{1}  D doublet in radial velocity scale relative to the systemic velocity, $RV_{\odot}$ = $-$40.8~km s$^{-1}$, are displayed in Figure~\ref{split}. All of the selected lines are resonance or low-excitation lines those cores are formed mainly in the outer layers of the stellar atmosphere. A medium strong \ion{Fe}{1} line at 6411.65 \AA\ with a Gaussian profile is shown for comparison purposes. As can be seen, the profiles of selected lines have a few components in the range of Doppler velocities from about $-$50 to +50~km s$^{-1}$ relative to the systemic velocity. The intensity and velocity of these components depends on  pulsation phase. Redshifted  components of variable intensity are seen at about 25~km s$^{-1}$ in the \ion{Na}{1} D12 and \ion{K}{1} profiles and at about 60~km s$^{-1}$ in the \ion{Na}{1} D12 profiles. Blue wings up to about 50~km s$^{-1}$ are clearly visible in the spectra  observed on October and November 2006. One of such broad and split absorption feature at 6390 \AA\ was analyzed in detail using quantitative methods (Figure~\ref{La2_profile}). Some weaker contributing atomic lines are identified and included in the list of lines for  synthesis  of the absorption profile to understand better the nature of splitting. Their positions are marked by vertical ticks according to the VALD3 database: \ion{Sm}{2} line at 6389.83 \AA,  \ion{Nd}{2} line at 6389.97 \AA,  \ion{La}{2} line at 6390.48 \AA, and \ion{Ce}{2} line at 6390.61 \AA. The profile at 6390.48 \AA\ was calculated with two final atmospheric models for  $T_\mathrm{eff}= 5750~ \mathrm{K}$ and 5250$~\mathrm{K}$; however, the synthesized profiles are not able to reproduce the observed blue/red-shifted components.  Thus, the sequence of profiles indicates probably that expanding and infalling layers are simultaneously present in the outer atmosphere of central star of IRAS\,22272+5435.  We examined  short-term variability in the profiles of split lines employing the time series of spectra observed near the light maximum (Sep 2010) and minimum (Oct 2011). An example of such an analysis is shown in Figure~\ref{La2_profile} (right panel) for an absorption feature formed mainly by the \ion{La}{2} line at 6390.48 \AA.  We detected a significant variation of intensity in the absorption component shifted to the red relative to the stellar photospheric velocity by about  10~km\,s$^{-1}$. A monotonic decrease of the intensity  for this component was observed over 12 days from September 17 to September 28. It should be noted that short-term variability was detected in the most of split low-excitation profiles examined here. The splitting and variability in the time scale of a week seem to be observational evidence of shock waves in the outer atmosphere of the star.

\begin{figure}
 \resizebox{\hsize}{!}{\includegraphics{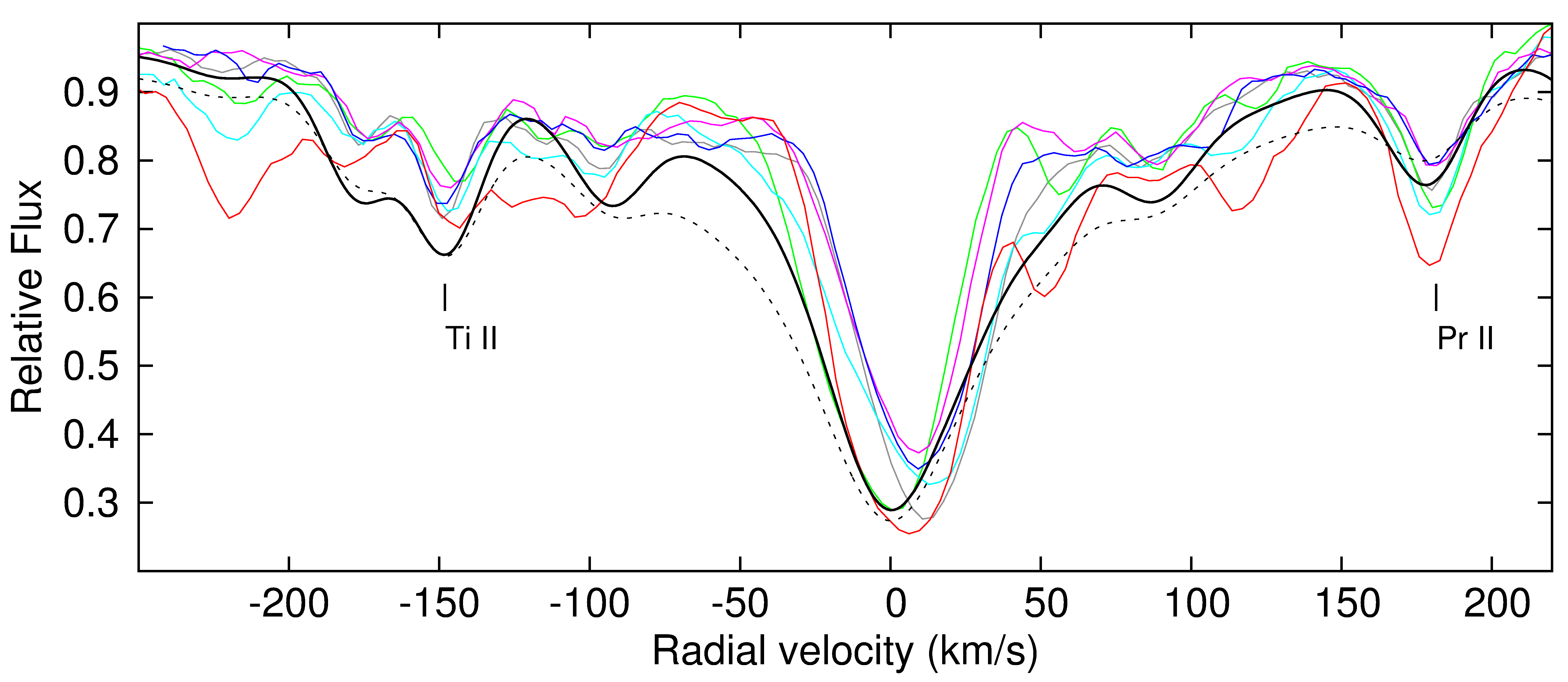}}
 \caption{The $H_{\alpha}$ profile observed on November 2002 (grey), October 2006 (cyan), November 2006 (green), February 2008 (magenta), September 2010 (blue), and November 2011 (red) in the radial velocity scale relative to the photosphere. The profiles calculated with SAM12 atmospheric models for  $\log g$ = 0.5 (cgs) and  for two effective temperatures, $T_\mathrm{eff}= 5750~ \mathrm{K}$ (black dashed) and 5250$~\mathrm{K}$ (black solid), are shown.}
   \label{Ha_RV_6}
\end{figure}

The profile of the Balmer $H_{\alpha}$ line is complicated and variable in the spectrum of IRAS\,22272+5435.  
The broad wings of the absorption profile are approximately in agreement with those calculated for supergiant with the atmospheric model for $T_\mathrm{eff}$ = 5250$~\mathrm{K}$.  However, the central part of the profile displays a shell-like emission with a central absorption (see Figure~\ref{Ha_RV_6}). Such profiles with a central absorption within the emission were found to be common for PPNe of spectral type F  \citep{hrivnak03}. Emission features of about equal height for IRAS\,22272+5435 are located on the both sides of the central absorption feature.  A time dependence in the profiles is seen in Figure~\ref{Ha_RV_6}, along with the profiles calculated with the two final models for $T_\mathrm{eff}= 5750~ \mathrm{K}$ and 5250$~\mathrm{K}$. The intensity and position of the central absorption feature is variable relative to the calculated profile. As can be seen, in the spectrum observed on 2006 November the position and intensity of the central absorption agree well with that calculated for the photosphere. In the other observed spectra the central absorption is redshifted relative to the photospheric velocity. Radial velocities measured for the central absorption in the spectra observed on November 2002, October 2006, November 2006, February 2008, September 2010, and November 2011 are, relative to the systemic velocity,  $\delta$RV = +7.5, +9.7, +3.8, +9.7, +8.9, +4.8~km\,s$^{-1}$, respectively. Thus, the central absorption is redshifted relative to the systemic velocity in all the observed spectra. The Balmer $H_{\beta}$ line is in absorption at approximately the photospheric velocity.

\subsection {$C_2$ Swan system lines}

Inspection of the observed high-resolution spectra revealed a significant intensity variation in the wavelength regions where the bandheads of $C_2$ Swan system are located. The spectra around Swan system bandhead (0,0) at 5165 \AA\ and the bandhead (0,1) at 5635 \AA\ were examined in details.  We varied the effective temperature  of atmospheric models and measured the radial velocity of selected lines to understand the reason of variation.

\begin{figure*}
  \resizebox{\hsize}{!}{\includegraphics{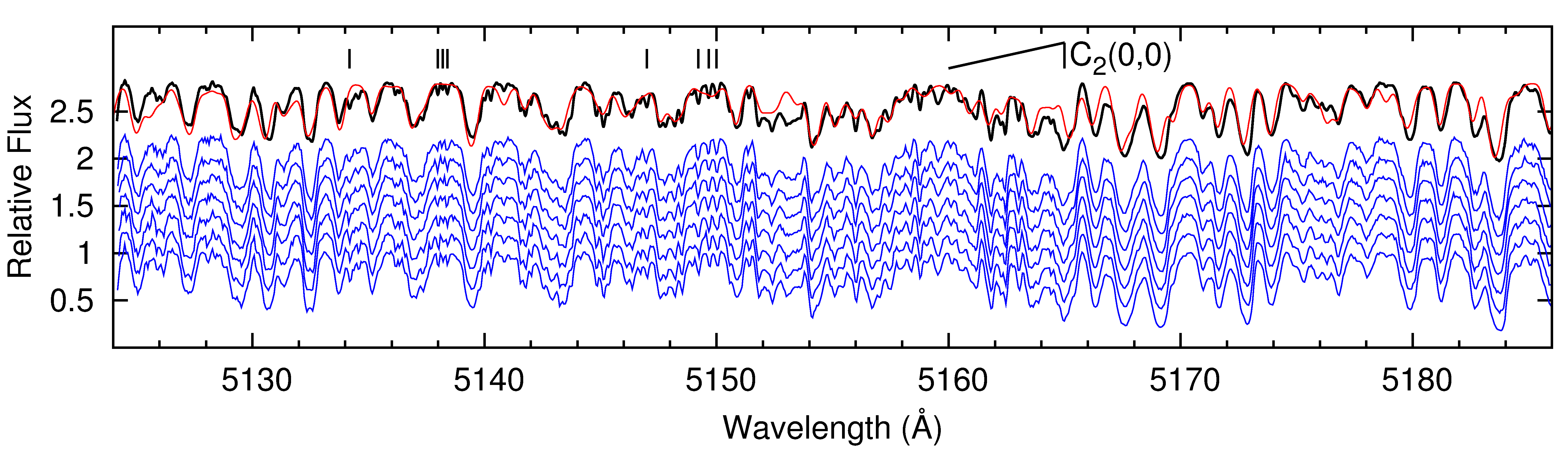}
   }
   \caption{Seven high-resolution spectra observed at the light maximum on 2010 September 17, 20, 21, 22, 25, 27, and 28  (blue) in the  wavelength region around $C_2$ Swan system (0,0) bandhead at 5165  \AA. The wavelengths are corrected for the instrumental photospheric radial velocity.  The upper spectrum (black) is a combined observed spectrum, along with a synthesized spectrum (red) calculated using the \citet{kurucz93} model for $T_\mathrm{eff}$ = 5750~K,  $\log g$ = 0.5 (cgs),  $\xi_t$ = 4.5 km s$^{-1}$, and abundances given in Section 3.2. Eight circumstellar $C_2$ lines given in the Table~\ref{CS_ID} are marked by vertical ticks on the top.}
  \label{C2_5165_2010}
 \end{figure*}

\begin{figure*}
 \resizebox{\hsize}{!}{\includegraphics{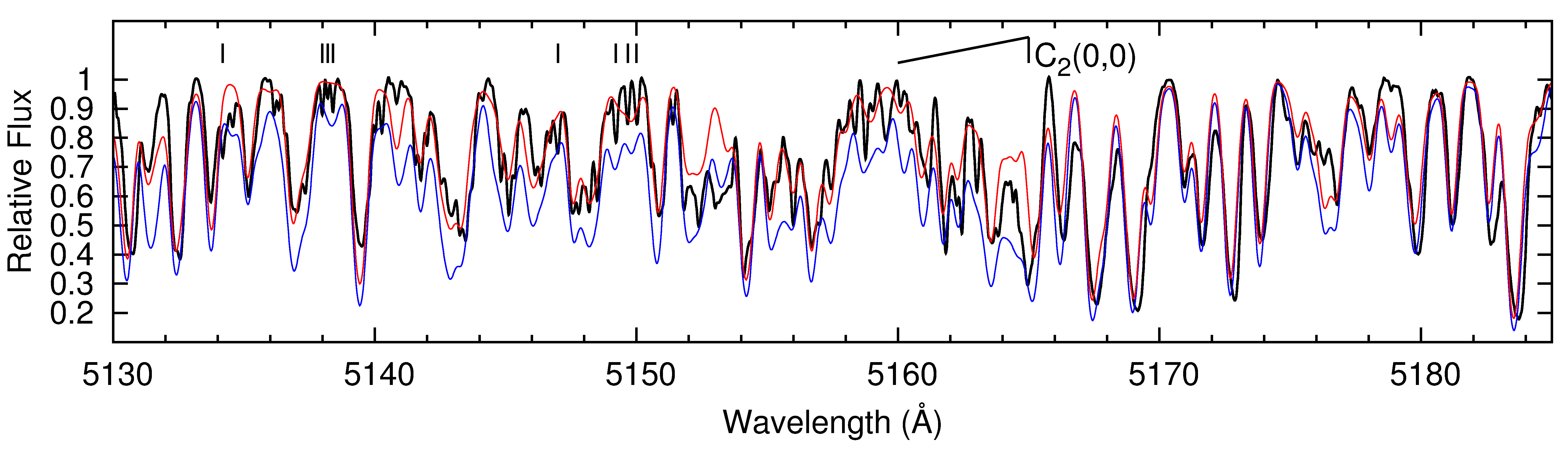}
  }
 \caption{Combined spectrum at the light maximum on September 2010 (black)  in the  wavelength region around $C_2$ Swan system (0,0) bandhead at 5165  \AA, along with two spectra calculated using SAM12 atmospheric models for $\log$g = 0.5 and two effective temperatures, $T_\mathrm{eff}$ = 5750~$\mathrm{K}$ (red) and 5250~$\mathrm{K}$  (blue).}    
\label{C2_5165_SAM12}
\end{figure*}

Seven high-resolution spectra in the wavelength region from 5125 to 5187 \AA\ gathered for IRAS\,22272+5435 during the set of observation from 2010 September 17  to September 28 at the light maximum (color index minimum) are displayed in the Figure~\ref{C2_5165_2010}. The spectra are shifted in the intensity scale for clarity. We concluded that there are no variations in the intensity and wavelength of the absorption lines among the individual spectra that are above the limit of detection in the this wavelength region.  Thus the resulting combined spectrum has a very high S/N ratio. This combined spectrum is given on the top of the Figure~\ref{C2_5165_2010}. The photospheric spectrum for the central star of IRAS\,22272+5435 was calculated with the standard Kurucz (1993) model for the atmospheric parameters ($T_{eff}$ = 5750~K, $\log$\,g = 0.5, [M] = --0.5) determined by Reddy et al. (2002) and for the atmospheric chemistry given in the Section 3.2. The agreement between the observed and synthesized spectra for the atomic lines is quite good, confirming  in general the correctness of the adopted atmospheric parameters and abundances. However, the intensity of absorption features around 5165 \AA\ was found to be larger in the observed spectrum in comparison with the calculated one. A weak $C_2$ band head in the spectrum observed at the light maximum was suspected. Redward of the $C_2$\,(0,0) band head we detected weak and narrow lines that were identified with single $C_2$ lines (Table~\ref{CS_ID}). Unfortunately, the intensity of $C_2$\,(0,0) band head  in  the calculated spectrum was found to be below the limit of detection rejecting the photospheric origin of the observed band head. However,  the bandhead could be stronger for the chemically peculiar atmospheric model. Therefore, the photospheric spectrum of IRAS\,22272+5435 was calculated with new self consistent SAM12 atmospheric models for two effective temperatures, $T_{eff}$ = 5750 (estimated temperature at maximum light) and 5250~K (estimated temperature at minimum light), with the goal  to check the impact of the adopted models on the synthesized spectrum and of helping to clarify the formation site of the detected $C_2$ features. For the hot SAM12 atmospheric model the intensity of band  head in the calculated spectrum is still below the limit of detection. The fit between observed and calculated line intensities for the SAM12 model with $T_{eff}$ = 5250~K is quite good near the $C_2$ bandhead, except for the position of the bandhead (Figure~\ref{C2_5165_SAM12}). However, the calculated  intensities of the $C_2$ photospheric lines blueward of the bandhead are too strong in the cool model and the profiles are broader in comparison with the observed ones. For these reasons we conclude that the site of formation of the weak $C_2$\,(0,0) band head and narrow single lines lies outside the photosphere of the central star of IRAS\,22272+5435.

\begin{table}
\caption[]{Blueshifted molecular lines in the spectra of IRAS22272+5435. 
The rest wavelengths, identifications, measured radial velocities relative to the systemic velocity, 
full widths at half maximum, and equivalent widths are given. \label{CS_ID}}
\begin{tabular}{cccccc}
\hline
\hline
\noalign{\smallskip}
 Wavelength &  Id   & $\delta$RV  & FWHM   &  EW     \\
   ( \AA\ ) &      & (km s$^{-1}$)  & ( \AA\ ) & (m\AA\ )   \\
\noalign{\smallskip}
\hline
\noalign{\smallskip}  
\noalign{\smallskip}
 Sep 2010  &    &      &    &    \\ 
\noalign{\smallskip}
  5134.31  & $C_2$ &   -8.8    &  ...     &  ...     \\
 5138.10  & $C_2$ &  -8.3    &  0.12  &  15     \\
 5138.31  &  $C_2$ &  -7.9   &  ...     & ...     \\
5138.50   &  $C_2$ &  -7.7   &   0.12 &  15    \\ 
5147.11   &   $C_2$ & -8.3   &   ...    &  ...    \\
5149.32   &  $C_2$ &  -7.6    &  ...     & ...     \\
5149.78   &  $C_2$ & -8.1   &  0.14  &  25    \\
5150.12   &  $C_2$ & -7.4   &   0.17 &  28    \\
5635.22:   & $C_2$ &  -16.8  &  0.49   &  41  \\
7865.88   &  $CN$  &  -8.9  & 0.29 & 25       \\
7868.66   &  $CN$  &  -8.8  & 0.25 &  27     \\ 
7871.65    & $CN$  &   -8.2  &  0.27 &  30    \\
\noalign{\smallskip}
Nov 2011  &    &   &  &   \\
\noalign{\smallskip}
5149.32    &  $C_2$ & -8.2  &  0.30   & 188   \\
5149.78    & $C_2$ & -7.1  &  0.18   &   65   \\
5150.12    & $C_2$ &  -7.3 &  0.24   & 104   \\
5629.14:  & $C_2$ &  -6.0    & 0.40  & 146   \\
5631.05:  & $C_2$ &  -5.4    & 0.44  & 165   \\
5635.22:  & $C_2$ &  -10.5  & ...     & ...      \\
6372.79    &  $CN$  &  -4.7   &  0.34   &   61   \\
6392.18    &  $CN$  &  -6.0   &   0.41  & 120   \\
6394.50    & $CN$  &   -6.1   &   0.28  &  40    \\
6396.37    & $CN$  &   -5.0   &   0.32  &  73    \\
6405.44    & $CN$  &   -5.4   &   0.34  &  88    \\
6415.41    & $CN$  &   -5.1   &   0.37  &  89    \\
6420.72    & $CN$  &   -6.5   &   0.42  & 120   \\
7865.88    & $CN$  &  -8.7  & 0.32     & 74    \\
7868.66    & $CN$  &  -8.9  & 0.32    &  76    \\
7871.65    & $CN$  &   -7.7 &  0.25  &  45    \\
  \noalign{\smallskip}
\hline
\end{tabular}
\end{table}

\begin{figure*}
\resizebox{\hsize}{!}{\includegraphics{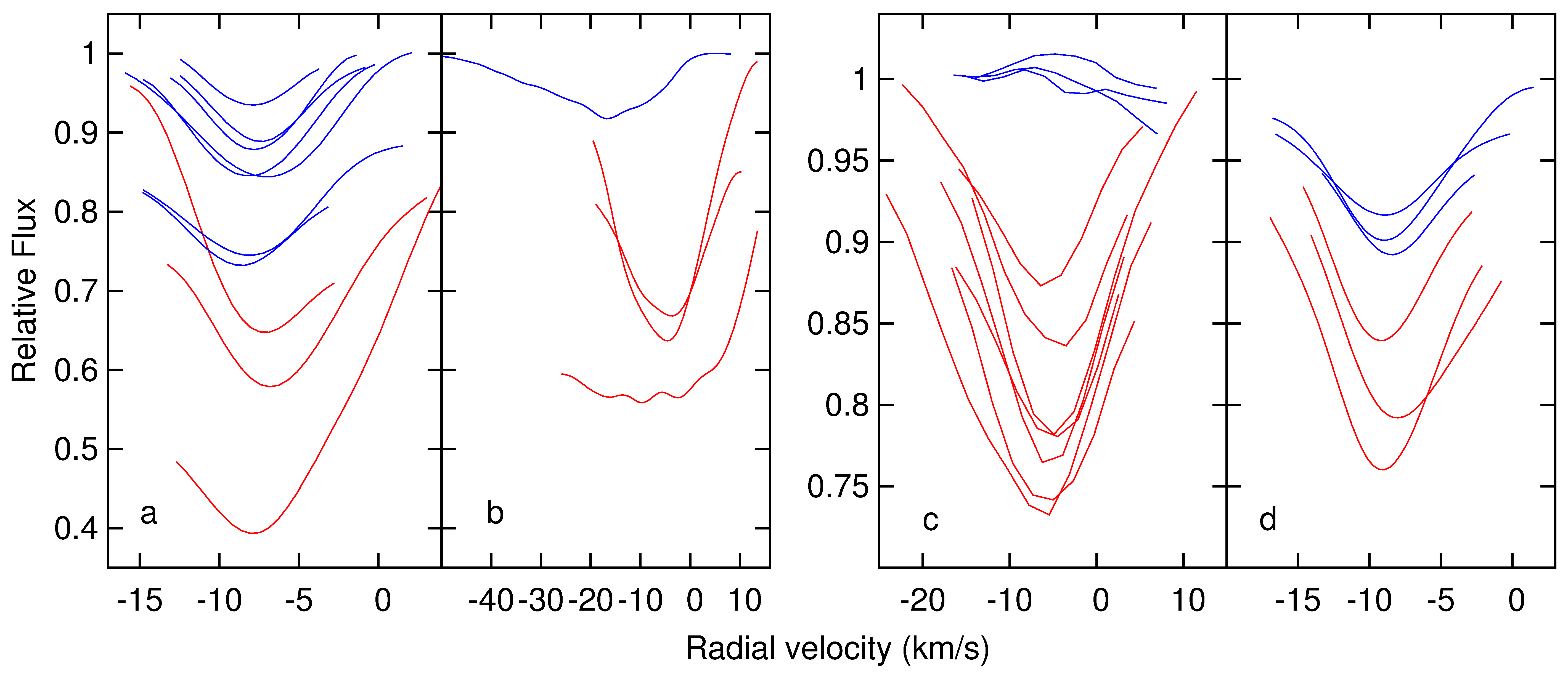}}
\caption{Blueshifted lines of carbon bearing molecules measured at light maximum on September  2010    
              (blue) and at light minimum on November 2011 (red) in the radial velocity scale relative to the systemic radial  
              velocity given in Table 2:  
               (a)  $C_2$\,(0,0) Swan system; 
               (b) $C_2$\,(0,1) Swan system; 
               (c) CN\,(5,1) Red system; 
               (d) CN\,(2,0) Red system.}
\label{CS}
\end{figure*}

The profiles of seven narrow $C_2$ lines shortward of the (0,0) bandhead observed at the light maximum on September 2010 are displayed in Figure~\ref{CS} (panel a) in radial velocity scaled relative to the systemic velocity (see Table~\ref{T_RV}).  
The lines are shifted  by $\delta$RV = --8.0 $\pm$ 0.4~km\,s$^{-1}$ relative to the systemic velocity. The blueshift and the width of these lines agree well with those observed for the circumstellar $C_2$  Phillips system lines in the spectrum of IRAS\,22272+5435,  $\delta$RV = --8.4 $\pm$ 0.5~km\,s$^{-1}$ \citep{zacs09}.  This would lead one to conclude that the site of formation of the narrow Swan system lines is the AGB shell.  However, in the spectra observed on November 2011 at the light minimum (color index maximum), the $C_2$\,(0,0) Swan system lines are much stronger.  Because most of them are blends at light minimum, we selected for quantitative comparison only three less blended $C_2$ lines  from the eight lines measured in the spectrum at the light maximum.  The measured radial velocities, FWHMs, and EWs are given in  Table~\ref{CS_ID} and the profiles are compared in Figure~\ref{CS} (panel a). As can be seen, the radial velocity of the blueshift $C_2$ lines relative to the systemic velocity is the same to within the 2$\sigma$ at the light maximum on September 2010 and at the light minimum on November 2011, $\delta$RV = --7.5$\pm$ 0.6~km\,s$^{-1}$, despite of the increased EWs and FWHMs at minimum light.

The spectra of IRAS\,22272+5435 obtained in different epochs and phases display large changes around the position of the $C_2$ Swan System (0,1) band head (Figure~\ref{C2_5635_5}).  The spectra observed near the light maximum on September 2010, on February 2008, and on November 2002  are similar and with a low molecular contribution. The band head of the $C_2$\,(0,1) Swan system is clearly seen, with an equivalent width of about 60 m\AA\ on average.  However, other lines of the Swan system are too weak to be identified with certainty. The bandhead is blueshifted by about 17~km s$^{-1}$ relative to the systemic velocity (Figure~\ref{CS}; panel b)  in the spectrum observed on September 2010 and its equivalent width changes from 87 to 60, and then 40 m\AA\ in the spectra observed near the light maximums on November 2002/February 2008/September 2010. In the spectrum observed at the light minimum on November 2011 the bandhead and the lines of $C_2$\,(0,1) system are extremely strong.  The measured blueshift of the bandhead is about 10~km s$^{-1}$ relative to the systemic velocity (lowest curve in  Figure~\ref{CS}, panel b).  Two less blended $C_2$ lines at 5629 and 5631 \AA\ give the blueshift of 6~km s$^{-1}$ on average relative to the systemic velocity for the line forming region.

\begin{figure*}
        \resizebox{\hsize}{!}{\includegraphics{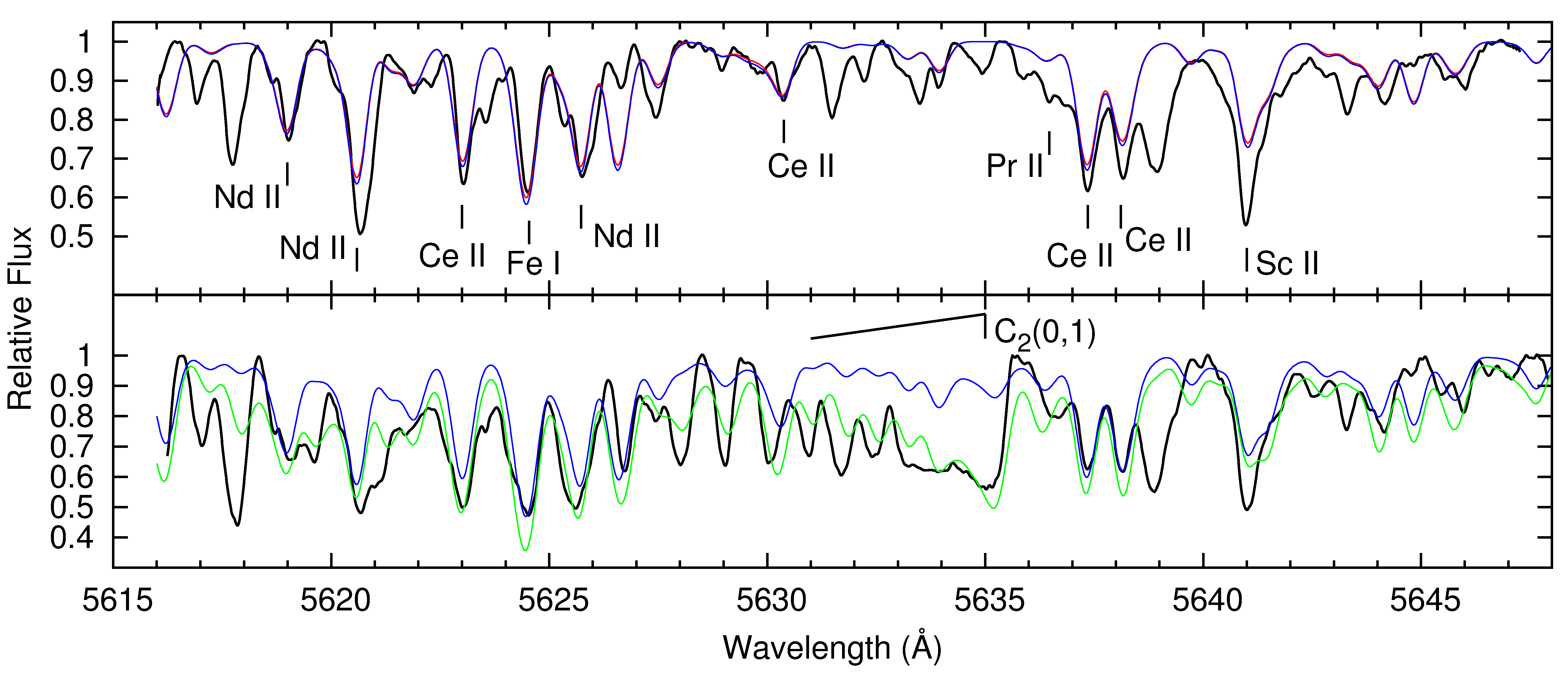}}
    \caption{The combined spectra at two different phases compared with modeled spectra in the wavelength region around the $C_2$(0,1) Swan system bandhead.  
    Upper panel: The spectrum observed at light maximum on September 2010 (black) along with two synthesized spectra calculated using the Kurucz's (red) and SAM12 (blue) models with $T_{eff}$ = 5750~K and $\log g$ = 0.5 (cgs).  
    Lower panel: The spectrum observed at the light minimum  on November 2011 (black), along with two synthesized spectra calculated using SAM12 models for  $\log g$ = 0.5 (cgs) and two effective temperatures, $T_{eff}$ = 5250~K (blue) and 4750~K (green).}
    \label{C2_5635_OBS_SYNT}
\end{figure*}

The spectrum around the $C_2$ Swan system (0,1) bandhead was synthesized using the \citet{kurucz93} and  SAM12 models, and the comparison is given in Figure~\ref{C2_5635_OBS_SYNT}. A fit between the spectrum observed at the light maximum and the spectra calculated with the hot atmospheric model for $T_{eff}$= 5750~K (estimated temperature at maximum light) is quite good for most of the identified s$-$process lines, indicating a correctness of the adopted  s$-$process abundances. However, the synthesized spectra have not reproduced the $C_2$ bandhead at 5635 \AA\ and a number of weak lines of unknown nature.  In addition, the synthesis of strong atomic lines is problematic, e.g. a poor fit for the feature at 5641 \AA\ identified with  the \ion{Sc}{2} line. The spectra calculated with the Kurucz's and the SAM12 models are similar, with the lines calculated with the SAM12 model being just a bit deeper. The spectrum observed at the light minimum on November 2011 was modeled with the cool  SAM12 model for  $T_{eff}$ = 5250~K.  In additional, the atmospheric model with $T_{eff}$ = 4750~K was employed to reproduce formally the observed intensity of the $C_2$ bandhead. 
As can be seen, the calculated photospheric spectra with $T_{eff}$ = 5250~K (estimated temperature at minimum light) and  4750~K have not reproduced the observed molecular spectrum at minimum light. The observed $C_2$ spectrum is blueshifted relative to the calculated ones and the molecular lines (e.g. at 5631.05 \AA) are more narrow in comparison with those in the modeled photospheric spectrum. 
This argues that the site of formation of the $C_2$ Swan system (0,1) bandhead is outside the photosphere. 
At the same time, one does see from the atmospheric models calculated for different temperatures in the current region that the variability observed in the atomic lines (e.g. \ion{Ce}{2} at 5637.36, 5638.11 \AA) is partly because of temperature variation in the atmosphere of star. 

\begin{table}
\caption[]{Selected atomic lines synthesized in the spectrum of HD\,235858.  Species identifications, the rest wavelengths, lower  excitation potentials, and oscilator strengths adopted from the VALD3 database are listed.  \label{atomic_lin}}
\begin{tabular}{cccccc}
\hline
\hline
\noalign{\smallskip}
 Species  & Wavelength &  LEP   &   $\log gf$ & Ref.   \\
               & ( \AA\ )        &  (eV)  &                 &             \\
\noalign{\smallskip}
\hline
\noalign{\smallskip}
\ion{Fe}{1} & 5166.28	&0.00	&-4.194  &  K07\\
\ion{Nd}{2} & 5170.94	&0.55	&-1.686 & MC \\
\ion{Fe}{1} & 5171.60	&1.48	&-1.792  & K07\\
\ion{Pr}{2}  & 5175.27	&0.22	&-1.241  & MC\\
\ion{Pr}{2}  & 5175.84	&0.42	&-1.282  & MC\\ 
\ion{Gd}{2} & 5176.29	&1.06	&-0.739  & DLSC \\
\ion{Nd}{2} & 5618.99	&1.77	&-0.649  & HLSC\\	
\ion{Nd}{2} & 5620.59	&1.54	&-0.309  & XSCL \\	
\ion{Ce}{2} & 5623.00        &0.96	&-1.150  & PQWB\\	
\ion{Ce}{2}  & 5623.03	&0.56	&-3.260 & PQWB \\	
\ion{Fe}{1} & 5624.54	&3.42	&-0.754  & K07 \\	
\ion{Nd}{2} & 5625.73	&0.93	&-1.119  & HLSC \\	
\ion{Ce}{2} & 5630.38	&1.64	&-0.940  & PQWB\\	
\ion{Pr}{2}  & 5636.47	&1.05	&-1.062  & BLQS \\	
\ion{Ce}{2} & 5637.36	&1.4	&-0.501  & PQWB \\	
\ion{Ce}{2} & 5638.11	&0.6	&-1.720  & PQWB \\	  
\ion{Pr}{2} & 5638.79	&0.63	&-1.070 & MC \\	
\ion{Sc}{2 }& 5641.00         &1.5	&-1.131 & LD \\	
\ion{Nd}{2} & 6382.06	&1.44	&-0.750  & HLSC \\	
\ion{Fe}{2} & 6383.72	&5.55	&-2.070  & BSScorr\\	
\ion{Nd}{2} & 6385.15	&1.16	&-0.770  & MC \\	
\ion{Nd}{2} & 6385.19	&1.6	&-0.360  & XSCL \\	
\ion{Sm}{2} & 6389.83       &1.17 &-1.959  & MC \\
\ion{Nd}{2} & 6389.97	&1.5	&-0.770  & HLSC \\	
\ion{La}{2} & 6390.48	&0.32	&-1.410  & LBS \\	
\ion{Ce}{2} & 6390.61         & 0.61 &-2.41   & PQWB \\
\ion{Fe}{1}& 6393.60          &2.43&-1.432  & K07 \\	
\ion{Pr}{2} & 6397.97	&1.05	&-0.940  & BLQS \\	
\ion{Fe}{1} & 6400.00          &3.6	&-0.289  & K07 \\	
\ion{Sm}{2} & 6406.25	&1.36	&-1.339  & LD-HS \\	
\ion{Fe}{1} & 6411.65	&3.65	&-0.594  & K07 \\	
\ion{Fe}{2} & 6416.92	&3.89	&-1.877  & BSScorr\\	
\ion{Fe}{1} & 6419.95	&4.73	&-0.240  & K07 \\	
\ion{Fe}{1}& 6421.35	&2.28	&-2.027  & K07 \\	
\ion{Ti}{2}& 6559.56	&2.05	&-2.175  & K10\\	
\ion{Pr}{2}& 6566.77	&0.22	&-1.721 & MC \\	
\ion{Ce}{2}& 7850.03	&0.74	&-1.289 & PQWB \\	
\ion{Ce}{2}& 7857.55	&0.9	&-1.289  & PQWB \\	
\ion{Ce}{2} & 7860.98	&0.7	&-2.560  & PQWB \\	
\ion{Y}{2} & 7881.88	           &1.84	&-1.569  & K11 \\	
\ion{Ce}{2}& 7898.97	&0.9	&-1.119  & PQWB \\	
\ion{Nd}{2}& 7900.39	&1.35	&-1.500  & MC \\	
\noalign{\smallskip}
\hline
\end{tabular}
\tablerefs{Ref. according to the VALD database:  \\ http://www.astro.uu.se/valdwiki/VALD3linelists}
\end{table}

\subsection {CN Red system lines}

\begin{figure*}
       \resizebox{\hsize}{!}{\includegraphics{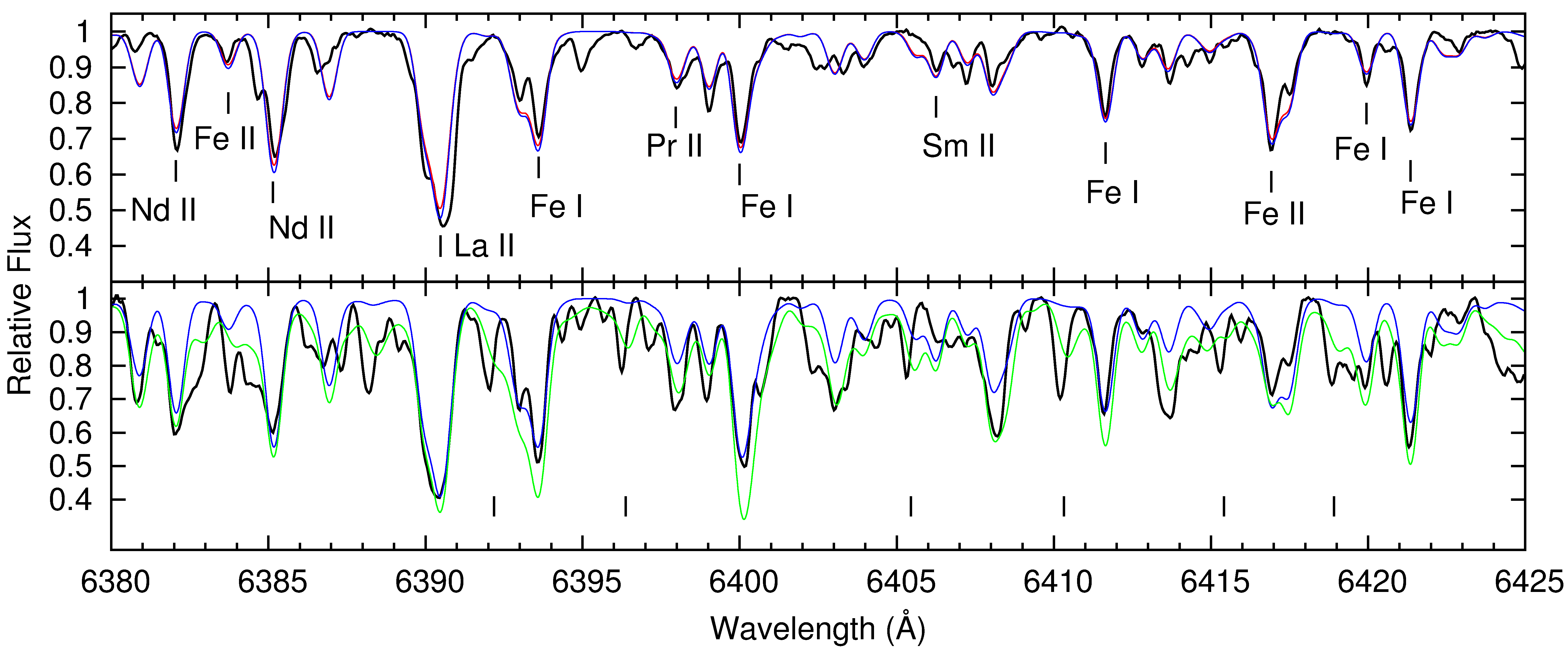}}
    \caption{Same as Fig.~\ref{C2_5635_OBS_SYNT}, displaying the spectra at light maximum (upper panel) and light minimum (lower panel), but in the wavelength region from 6380 to 6425 \AA. Some less blended lines of CN Red system (5,1) given in Table~\ref{CS_ID} are marked by tick marks.}
    \label{CN_51_OBS_SYNT}
\end{figure*}

A comparison of the observed spectra revealed a significant intensity variation of lines in the wavelength region 
around 6400 \AA\ (Figure~\ref{CN_51_5}). Most of the variability was found to be in the lines of the CN Red system. We varied atmospheric models and measured radial velocity of selected lines again  to understand the reason of variation. The spectrum observed at the light maximum on September 2010 was synthesized with the hot atmospheric model for $T_{eff}$= 5750~K (estimated temperature at maximum light). The fit among the calculated and the observed spectra is given in Figure~\ref{CN_51_OBS_SYNT}. 
We conclude that the intensity of CN Red system lines as calculated with the hot \citet{kurucz93} and SAM12 models are to weak to be identified in the spectra. In the spectrum observed on September 2010, weak emissions are suspected at the positions of CN\,(5,1) lines. We examined the presence of such emissions in the rest spectra and confirmed it in the spectra observed at light maximum on February 2008 and November 2002. On the other hand, a good fit between observed and calculated spectra was found for the s$-$process elements (Nd, La, Pr, Sm) and for the neutral and ionized iron lines (see list in Table~\ref{atomic_lin}), confirming the correctness of the adopted gravity, metallicity, and abundances of the selected s$-$process elements. The fit between the spectrum observed at the light minimum on November 2011 and the synthesized spectra is shown in Figure~\ref{CN_51_OBS_SYNT} (bottom panel). 
We concluded, that the  synthesized spectrum calculated with the cool atmospheric model for $T_{eff}$ = 5250~K (estimated temperature at minimum light) reproduce correctly the atomic lines, however, the modeled molecular lines are too weak in comparison with the observed ones.
At the light minimum, CN Red system lines are strong and there is significant blending with the atomic lines in this region. We concluded, that the  synthesized spectrum calculated with the cool atmospheric model for $T_{eff}$ = 5250~K reproduces correctly the atomic lines but the modeled molecular lines are too weak in comparison with the observed ones.
Even the spectrum calculated with the extremely cool model for $T_{eff}$ = 4750~K does not produce strong molecular lines. In addition, the observed molecular lines are more narrow than the calculated photospheric lines and they are blueshifted relative to the systemic velocity (Table~\ref{CS_ID}). The radial velocity measurements for selected CN Red system lines confirmed a  blueshift of the site where the CN\,(5,1) lines are formed at light minimum,  $\delta$RV = --5.5 $\pm 0.7~km\,s^{-1}$ (Figure~\ref{CS}, panel c).

\begin{figure*}
        \resizebox{\hsize}{!}{\includegraphics{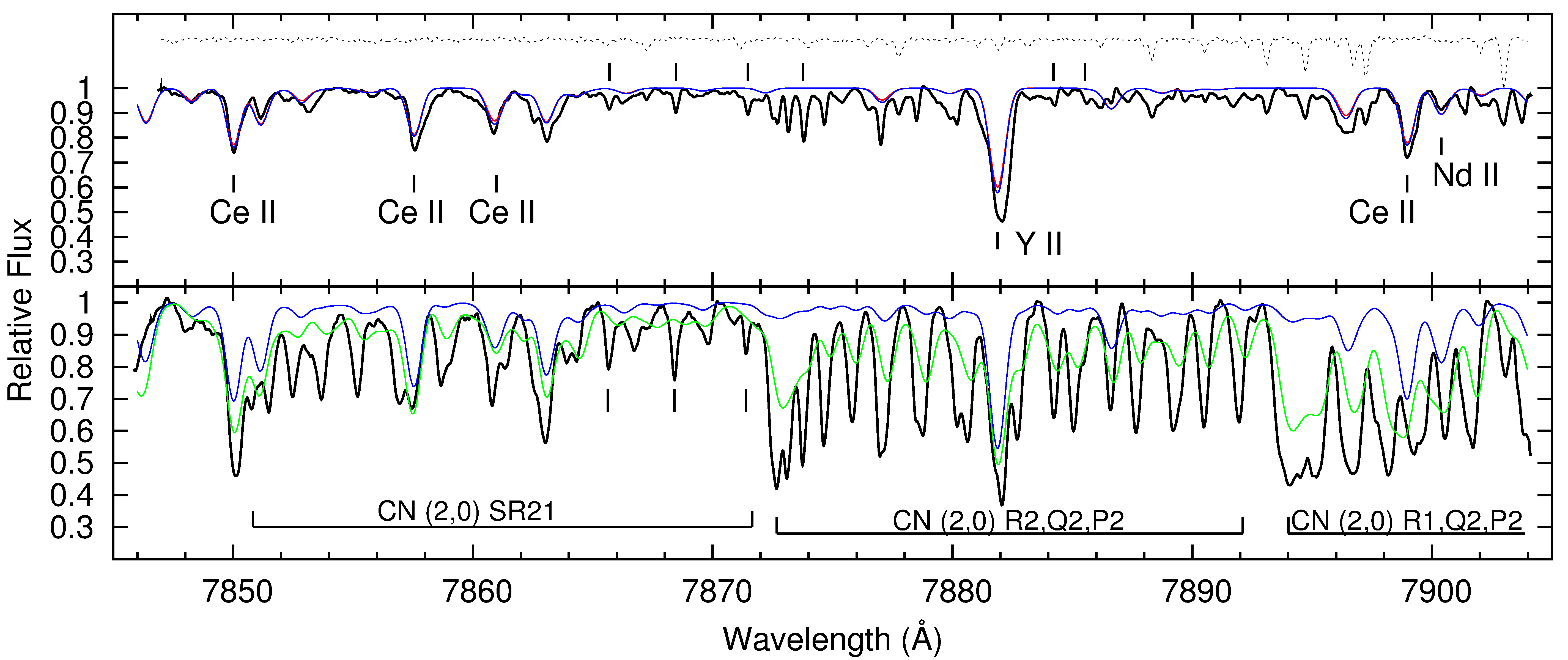}}
    \caption{Same as Fig.~\ref{C2_5635_OBS_SYNT}, displaying the spectra at light maximum (upper panel) and light minimum (lower panel), but in the wavelength region from 7845 to 7904 \AA. The telluric spectrum is shown by dotted line on the top. Six unblended lines of CN Red system (2,0) at light maximum and three less blended ones at light minimum are marked by tick marks. The measured and compared CN  lines are  listed in Table~\ref{CS_ID}.}
    \label{CN_20_OBS_SYNT}
\end{figure*}

A comparison of the observed spectra for IRAS\,22272+5435 in the near-infrared region around 7900 \AA~ revealed a dramatical intensity variation.  In Figure~\ref{CN_20_OBS_SYNT}  is displayed the spectra observed at the light maximum on September 2010 and at the light minimum on November 2011, along with a telluric spectrum observed simultaneously (upper panel). The atomic lines (see Table~\ref{atomic_lin}) and the narrow CN Red System (2,0) lines (see Table~\ref{CS_ID})  were identified in the spectrum observed at light maximum.  A fit between the observed spectrum and that calculated for the hot model with $T_{eff}$ = 5750~K (estimated temperature at maximum light) confirms  in general the accepted s$-$process abundances and some previously noted problems in fitting the profiles for strong atomic lines (Figure~\ref{CN_20_OBS_SYNT}). We see that photospheric CN lines are too weak in the synthetic spectrum calculated with the hot atmospheric model to expect them to be visible in the photosphere of the star. Thus their presence points to a non-photospheric formation region.  The width and shift of the identified CN\,(2,0) lines confirm their circumstellar origin, with $\delta$RV = --8.6 $\pm 0.4~km\,s^{-1}$  (Figure~\ref{CS}; panel d). 
The CN\,(2,0) Red system lines are extremely strong in the spectrum observed at the light minimum on November 2011 and the atomic lines are significantly blended in that wavelength region. The spectra calculated with the cool atmospheric model for $T_{eff}$ = 5250~K (estimated temperature at minimum light) and with the extra cool atmospheric model for $T_{eff}$ = 4750 ~K have not reproduced the observed spectrum -- the observed CN lines are much stronger and narrower in comparison with the modeled photospheric lines. The radial velocity measured in the spectrum observed on November 2011, using the three less blended CN lines, confirms a blueshifted of the line forming site, with
$\delta$RV = --8.4 $\pm 0.6~km\,s^{-1}$, relative to the systemic velocity (Table~\ref{CS_ID}). Thus, the blueshift of the CN lines relative to the systemic velocity is the same within 1$\sigma$ at the light maximum and at the light minimum, despite of the large changes in their EWs and FWHMs.

\section{Discussion and conclusions}

A time series of high-resolution spectra  in the optical wavelength region was collected and analyzed  for the cool  proto-planetary nebula IRAS\,22272+5435. In total seventeen spectra were observed between 2002 November and 2011 November and the phase of pulsation was estimated for each spectrum based upon  simultaneous light and radial velocity monitoring. Inspection of the spectra revealed a significant intensity variation in the molecular and atomic lines.  The spectral variations  could be an objective reason of the contradictory spectral classifications and abundances published for the central star \citep{skiff, zacs95, reddy02}.                

The $C_2$ Swan system and the CN Red system lines are stronger near the light minimum when the star is cooler according to the observed color indices. The effective temperature was accepted to be $T_{eff}$ = 5750~K at maximum light  for the central star and the probable variation of the effective temperature due to the pulsation was estimated using the color index ($V-R_C$), $\delta\,T_{eff} \simeq$500~K. 
A grid of self consistent hydrostatic atmospheric models  was calculated for a range of effective temperatures to reproduce approximately the maximum changes of the atmospheric structure for the central star over the pulsation cycle. The photospheric spectrum of the central star was synthesized using models for the effective temperatures  between 4750 and 5750~K, and these modeled spectra were compared with the observed spectra to help clarify the  reasons for the spectroscopic variability. The observed intensity variations in $C_2$ Swan system and CN Red system lines were found to be much larger than could be due to the temperature variation in the atmosphere of the central star. In addition, the observed molecular lines are  more narrow in comparison with the photospheric atomic lines and  they are blueshifted relative to the photospheric velocity. Thus, the site of the formation of molecular lines should be outside the atmosphere of the central star. The intensity variations in the atomic lines seem to be mostly due to variations in the effective temperature during the pulsation cycle. However, the profiles of the strong atomic lines bear evidence of shock waves in the atmosphere of IRAS\,22272+5435.

The profiles of strong low-excitation atomic lines were found to be split into two or more components.  The intensity and radial velocity of these components depends upon the pulsation phase. The velocities range between about $-$50 and +50~km s$^{-1}$ with the most common blueshifts and redshift of about 10~km s$^{-1}$. The profiles calculated using the hydrostatic atmospheric models were not able to reproduce the observed multi-component features. Moreover, intensity variations in a timescale of days were observed in the split profiles. It is generally accepted that complex line profiles in the spectra of long-period variables are related to the shock wave and the associated velocity gradient in the atmosphere of the pulsating star. The occurrence of shock fronts for the Mira variables has been suspected from observations of emission lines around the light maximum \citep{richter} and from splitting of the profiles of molecular and atomic lines \citep{hinkle82, alvarez1}. The role of velocity fields in the formation of split absorption lines has been confirmed thanks to the cross-correlation technique and monitoring of Miras \citep{alvarez1}.  \citet{alvarez2} observed the line-splitting phenomena for 54 \% stars in the sample of 81 long-period variables (LPV) of various  periods, spectral types, and brightness ranges, and most of them display emission in the Balmer lines. The measured velocity differences between the blue and red peaks for LPVs range from 10 to 25~km s$^{-1}$. The velocity of the shock front is a matter of debate, because a large shock-wave velocity of the order of 60~km s$^{-1}$ is required to ionize the hydrogen atoms \citep{gillet, alvarez2}. \citet{nowotny1, nowotny2} studied the atmospheric dynamics of carbon-rich Miras using model atmospheres and synthetic line profiles and confirmed a splitting phenomena in the absorption lines. The models confirmed propagation of the shock (compression) waves throughout the stellar atmosphere, with the outward and downward waves reaching amplitudes of about 20 and  10--15~km s$^{-1}$, respectively. Thus,  the observed splitting and short-term variability in low-excitation atomic lines provide strong evidence that shock waves are present in the outer layers of the atmosphere of the central star IRAS\,22272+5435. The variable emission features inside the shell-like profile of the Balmer $H_{\alpha}$ line could be one more indication of the propagation of shock waves through the photosphere of this pulsating star \citep{lebre}. The lack of line splitting phenomena in the high excitation-potential lines is most likely because these lines are formed deeper in the atmosphere. 

The appearance of strong and variable absorption lines of carbon-bearing molecules in the spectrum of IRAS\,22272+5435 raised the question about their site of formation. The $C_2$ Swan system band heads and the CN (2,0) Red system lines are weak in the spectra observed at the light maximum. The photospheric spectrum calculated with the atmospheric model suitable for the central star at the light maximum ($T_\mathrm{eff}$ = 5750~$\mathrm{K}$), when the star is hottest, has not reproduced the observed intensity of the band heads --  the model for  lower effective temperature is needed.  Photospheric lines of the CN Red system calculated with the model for $T_\mathrm{eff}$ = 5750~$\mathrm{K}$ are below the level of detection. The lines of the CN Red system (5,1) are suspected to be in a weak emission at the light maximum on November 2002, February 2008, and September 2010. Weak and narrow $C_2$ Swan system (0,0) and CN Red system (2,0) lines identified in the spectra at the light maximum on September 2010 are blueshifted  by $-$8.3~km s$^{-1}$  relative to the systemic velocity, and are attributed to formation in the AGB shell with the expansion velocity in range between 7.5 to 9.0~km s$^{-1}$  estimated using CO emission lines by \citet{hrivnak05}. The site of formation of the $C_2$  (0,0) band head at light maximum, which is blueshifted relative to the systemic velocity by about 17~km s$^{-1}$,  is uncertain. 

The $C_2$ Swan system band heads and lines of the system (0,1) are extremely strong and blueshifted relative to the systemic velocity in the spectra observed at light minimum. The spectrum calculated with the atmospheric model for effective temperature suitable for the central star of  IRAS\,22272+5435  at the light minimum ($T_\mathrm{eff}$ = 5250~$\mathrm{K}$), when the star is the coolest,  has not reproduced the intensity of the band heads -- a lower effective temperature is needed. A blueshift measured using less blended $C_2$ lines was found to be $\delta$RV =  --7.0~km s$^{-1}$ relative to the systemic velocity, which is slightly lower in comparison with that measured at the light maximum; However, agrees still within uncertainty with  the expansion velocity of the AGB shell. Notice that the expansion velocity of the AGB remnant estimated using a number of $C_2$ and CN absorption lines by \citet{bakker} and \citet{reddy02}  ranges between 6.5 to 11.4~km s$^{-1}$ relative to the the  systemic velocity accepted in this paper, $RV_{sys}^{\sun}$ = -40.8~$km\,s^{-1}$. Notice, too, that \citet{nakashima} introduced two circumstellar expanding structures for their IRAS\,22272+5435 model with the  expansion velocities of 7.5 and 10.5~$km\,s^{-1}$, respectively. Thus, the strong $C_2$ lines observed at light minimum could be attributed to formation in the AGB ejecta according to the measured blueshift.  However, the observed intensity variation in time  and phase dependence reject to attribute them to formation in the AGB shell which is located at the distance of about $10^{16}$ cm from the star \citep{ueta2, nakashima}. We concluded instead that the strong $C_2$ features observed at light minimum were formed in a cool outflow triggered by pulsation. For the measured CN lines our conclusion is similar. The CN Red system lines at light minimum are strong and the spectra calculated with the cool atmospheric model ($T_\mathrm{eff}$ = 5250~$\mathrm{K}$) and with the extra cool model ($T_\mathrm{eff}$ = 4750~$\mathrm{K}$) have  not reproduced the intensity of the observed lines. In addition, the observed CN lines are more narrow than would be expected in the atmosphere of the star and they are blueshifted. The blueshift value for the CN lines is the same within uncertainties at both the light minimum and at the light maximum. The strong and blueshifted CN lines are apparently formed near the star, e.g. in the cool outflow, with an  expansion velocity similar to that measured for the AGB ejecta of IRAS\,22272+5435.  Thus the molecular lines which are formed in the pulsationally-triggered cool outflow and those formed in the AGB shell are overlapping. Spectra of higher resolution are needed to confirm such an interpretation.

\acknowledgments
 L.Z. thanks staffs of Valparaiso University, Vilnius University Observatory, and Main Astronomical Observatory of Ukraine for hospitality during stay on April/May 2013, August/September 2014, and March 2015, when the current paper was designed and prepared. We acknowledge support for this collaboration from the EU  FP7-PEOPLE-2010-IRSES program  in the framework of project POSTAGBinGALAXIES (grant agreement No. 269193).  L.Z. and J.S. acknowledge support from the Research Council of Lithuania under the grant MIP-85/2012.  B.J.H. acknowledges support from the National Science Foundation (AST 1009974, 1413660). 
This research has made use of the Simbad database operated at CDS, Strasbourg, France, and the VALD database, operated at Uppsala University, the Institute of Astronomy RAS in Moscow, and the University of Vienna.

\end{document}